\documentstyle[12pt]{article}
\renewcommand{\theequation}{\thesection.\arabic{equation}}
\newlength{\extraspace}
\newlength{\extraspaces}
\newcommand{\be}{\begin{equation}
\addtolength{\abovedisplayskip}{\extraspaces}
\addtolength{\belowdisplayskip}{\extraspaces}
\addtolength{\abovedisplayshortskip}{\extraspace}
\addtolength{\belowdisplayshortskip}{\extraspace}}
\newcommand{\ee}{\end{equation}}
 
\newcommand{\ba}{\begin{eqnarray}
\addtolength{\abovedisplayskip}{\extraspaces}
\addtolength{\belowdisplayskip}{\extraspaces}
\addtolength{\abovedisplayshortskip}{\extraspace}
\addtolength{\belowdisplayshortskip}{\extraspace}}
\newcommand{\ea}{\end{eqnarray}}
 
\newcommand{\bas}{\begin{eqnarray*}
\addtolength{\abovedisplayskip}{\extraspaces}
\addtolength{\belowdisplayskip}{\extraspaces}
\addtolength{\abovedisplayshortskip}{\extraspace}
\addtolength{\belowdisplayshortskip}{\extraspace}}
\newcommand{\eas}{\end{eqnarray*}}
 
\newcounter{subequation}[equation]
\makeatletter
\expandafter\let\expandafter\reset@font\csname reset@font\endcsname
\def\subeqnarray{\arraycolsep1pt
    \def\@eqnnum\stepcounter##1{\stepcounter{subequation}
        {\reset@font\rm(\theequation\alph{subequation})}}\eqnarray}

\makeatother
 
\newenvironment{theorem}[1]
{\vspace{3mm}\noindent {\bf #1 :} }{\vspace{2mm}}
\newcommand{\bt}[1]{\begin{theorem}{#1}}
\newcommand{\et}{\end{theorem}}
 
\newcommand{\newsection}[1]{
\vspace{12mm}
\pagebreak[3]
\addtocounter{section}{1}
\setcounter{equation}{0}
\setcounter{subsection}{0}
 
\begin{flushleft}
{\large\bf \thesection. #1}
\end{flushleft}
\nopagebreak
\medskip
\nopagebreak}
 
\newcommand{\newsubsection}[1]{
\vspace{1cm}
\pagebreak[3]
 
\addtocounter{subsection}{1}
\noindent{ \bf \thesubsection. #1}
\nopagebreak
\vspace{2mm}
\nopagebreak}
 
\newcommand{\is}{\! & \! = \! & \!}

\newcommand{\pa}{{\partial}}

\newcommand{\cW}{{\cal W}}
 
\newcommand{\ps}{p\llap{/}} 
\newcommand{\Ds}{D\llap{/}}
\newcommand{\intd}{\int \! d^4 x \;}
\newcommand{\brs}{\mathrm s}
\newcommand{\cw}{\cos \theta_W}
\newcommand{\cws}{\cos^2 \theta_W}
\newcommand{\sw}{\sin \theta_W}
\newcommand{\sws}{\sin^2 \theta_W}
\newcommand{\cg}{\cos \theta_G}
\newcommand{\sg}{\sin \theta_G}
\newcommand{\cwg}{\cos (\theta_W- \theta_G)}
\newcommand{\swg}{\sin (\theta_W- \theta_G)}

\newcommand{\fsc}{{e^2 \over 16 \pi ^2}}

\def\a{\alpha}
\def\be{\beta}
\def\ga{\gamma}

\def\Ga{\Gamma}

\def\refeq#1{\mbox{(\ref{#1})}}

\def\citere#1{\mbox{Ref.~\cite{#1}}}
\def\citeres#1{\mbox{Refs.~\cite{#1}}}

\def\draftdate{\relax}
\def\mda{\relax}
\def\mua{\relax}
\def\mla{\relax}
\def\draft{
\def\thtystars{******************************}
\def\sixtystars{\thtystars\thtystars}
\typeout{}
\typeout{\sixtystars**}
\typeout{* Draft mode!
         For final version remove \protect\draft\space in source file *}
\typeout{\sixtystars**}
\typeout{}
\def\draftdate{\today}
\def\mua{\marginpar[\boldmath\hfil$\uparrow$]%
                   {\boldmath$\uparrow$\hfil}%
                    \typeout{marginpar: $\uparrow$}\ignorespaces}
\def\mda{\marginpar[\boldmath\hfil$\downarrow$]%
                   {\boldmath$\downarrow$\hfil}%
                    \typeout{marginpar: $\downarrow$}\ignorespaces}
\def\mla{\marginpar[\boldmath\hfil$\rightarrow$]%
                   {\boldmath$\leftarrow $\hfil}%
                    \typeout{marginpar: $\leftrightarrow$}\ignorespaces}
\def\Mua{\marginpar[\boldmath\hfil$\Uparrow$]%
                   {\boldmath$\Uparrow$\hfil}%
                    \typeout{marginpar: $\Uparrow$}\ignorespaces}
\def\Mda{\marginpar[\boldmath\hfil$\Downarrow$]%
                   {\boldmath$\Downarrow$\hfil}%
                    \typeout{marginpar: $\Downarrow$}\ignorespaces}
\def\Mla{\marginpar[\boldmath\hfil$\Rightarrow$]%
                   {\boldmath$\Leftarrow $\hfil}%
                    \typeout{marginpar: $\Leftrightarrow$}\ignorespaces}
\overfullrule 5pt
\oddsidemargin -2mm
\marginparwidth 29mm
}

\begin{document}
%
\begin{titlepage}
%
\renewcommand{\thefootnote}{\fnsymbol{footnote}}
\begin{flushright}
BN-TH-98-16\\
KA-TP-12-1998\\
hep-ph/9807548
\end{flushright}
\vspace{1cm}
 
\begin{center}
{\Large {\bf The Callan-Symanzik equation of the electroweak}} \\[4mm]
{\Large {\bf Standard Model and its 1-loop functions}}
{\makebox[1cm]{  }       \\[1.5cm]
{\bf Elisabeth Kraus}\\ [3mm]
{\small\sl Physikalisches Institut, Universit\"at Bonn} \\
{\small\sl Nu\ss allee 12, D--53115 Bonn, Germany} \\[0.5cm]
{and}\\[0.5cm]
{\bf Georg Weiglein }\\ [3mm]
{\small\sl Institut f\"ur Theoretische Physik, Universit\"at
Karlsruhe,} \\
{\small\sl Postfach 6980, D--76128 Karlsruhe, Germany}} 
\vspace{1.5cm}
 
{\bf Abstract}
\end{center}
\begin{quote}
We derive the Callan-Symanzik equation of the electroweak
Standard Model in the QED-like on-shell parameterization. The
various coefficient functions, the $\beta$-functions and anomalous
dimensions, are determined in one-loop order in the most general linear
gauge compatible with rigid symmetry. In this way the basic 
elements for a systematic investigation of higher-order leading
logarithmic contributions in the Standard Model are provided.
The one-loop $\beta$-function of the electromagnetic coupling turns out
to be independent of mass ratios and it is QED-like in this sense.
Besides the QED-contributions of fermions it contains non-abelian
contributions from vectors and ghosts with negative sign, which
overcompensate the contributions of the fermions if one restricts the
latter to one fermion generation.
We also compare our results with the symmetric theory and give 
relations between the $\beta$-functions of the spontaneously broken and
the symmetric theory valid in one-loop order.

\end{quote}
\vfill
\renewcommand{\thefootnote}{\arabic{footnote}}
\setcounter{footnote}{0}
\end{titlepage}
%

\newsection{Introduction}

The precision tests of the electroweak theory have reached such a high
level of experimental accuracy~\cite{datasum96} that in the perturbative
evaluation of the theoretical predictions the incorporation of 
higher-order radiative corrections is indispensable (for a recent review
see e.g.\ \citere{YeRep}). The theoretical predictions obtained in a fixed
order of perturbation theory can be improved if it is known how
the leading contributions can consistently be resummed to all orders.
In this context one is often also interested in the large-momentum
behavior of the contributing Green functions.
In order to make the analysis
of the large-momentum behavior meaningful, also contributions which
depend on large ratios of different mass scales
have to be considered.

Information of this kind can be obtained by studying 
the Callan-Symanzik (CS) equation~\cite{CAL70}
and the renormalization
group (RG) 
\cite{STPE53}
equation of the model under consideration. The CS equation describes
the breaking of dilatations and contains information about the momentum
structure of the theory. The RG equation on the other hand describes
the invariance of the model under variations of the normalization point.
Both equations can be systematically constructed in renormalized perturbation
theory. Their importance is founded in the fact that RG invariance
as well as the hard breaking of dilatations can be formulated as
a partial differential equation. They both contain derivatives with respect
to the independent parameters of the theory, which give rise
to the CS and RG $\beta$-functions, and field differential
operators, which are connected with the anomalous dimensions.

While these equations coincide in
theories with unbroken symmetry and massless particles, in massive
theories this is no longer the case. In theories with unbroken symmetry
the large-momentum behavior can be related to the large-momentum behavior
of the massless theory. In particular
it can be shown that for asymptotic normalization conditions the 
$\beta$-functions and anomalous dimensions of the massive theory
coincide with those of the massless theory~\cite{asnorm}.
 This situation is changed drastically in theories with broken
symmetries: In the physical on-shell schemes 
the CS equation has a different
form than the RG equation, in particular it contains derivatives with
respect to the physical masses of the theory. Moreover, solving both equations
consistently it has been shown that the massless symmetric theory is not
the asymptotic version of the spontaneously broken one, but contains
large logarithms of the mass parameters of the broken theory~\cite{InvCh}.
Consequently it is not obvious how 
to interpret the solutions of the CS equation in terms of ``running''
couplings and masses.
It is therefore not guaranteed that the results obtained from a
RG-study using the symmetric parameterization of the theory 
(cf.~\citere{Ara} and Refs.~therein and also \cite{KUSI85}) are directly
applicable to the Standard Model~(SM) of electroweak interactions. 
Instead, modifications are to be expected beyond one-loop order. 

As a first step towards a systematic
analysis of large-momentum {\it and\/} mass-dependent higher-order 
contributions,
we derive in the present paper the CS equation of the electroweak SM
in the on-shell parameterization 
(see e.g.~\citeres{onshell, bhs, Dehab})
and determine its 1-loop coefficient functions. 
The benefits of working within an on-shell parameterization are founded 
not only in its transparency due to the formulation in terms of physical 
parameters,
appropriate on-shell conditions for the mixing propagators involving
massless particles are also important for ensuring decent infrared properties
of higher-order Green functions.
We evaluate the coefficient functions of the CS equation
in the most general linear gauge
compatible with rigid symmetry, providing them in this way in the form
needed for higher-order investigations.
As an explicit example, the quadratic logarithms in the asymptotic
region are determined for the photon self-energy at two-loop order.
We also compare our results to the symmetric theory and to QED of
charged fermions. Concerning QED we find that the fermion 
contributions to the $\beta$-function of the electromagnetic coupling
 are the same as
in the electroweak SM. However, the non-abelian
contributions of ghosts and vectors enter with a negative sign.
In particular it turns out that if one restricts the
fermions of the SM to one generation, the one-loop $\beta$-function of the
electromagnetic coupling has a different sign compared to the familiar
QED $\beta$-function. In this context it should be noted that in QED
there exist also partial differential equations with respect to
variations of single fermion masses. Such equations cannot be derived
in the electroweak SM.
This gives rise to the fact 
that higher-order $\beta$-functions are not restricted from abstract
analysis in their mass-parameter dependence.

Apart from the above-mentioned applications
the CS equation is also an important object in the procedure
of abstract renormalization.
 It allows to determine in a scheme-independent
way the independent parameters of the theory.\footnote{For an introduction
 to algebraic renormalization see
 \citere{PISO95}.}
The most important outcome of the present analysis in this context
is the observation that the ghost mass ratio is an independent parameter
of the model, i.e.~it is renormalized independently from the vector
mass ratio. In order to introduce it as an independent parameter
we have to modify  the BRS transformations in lowest order.
Otherwise it is not possible to assign 
well-defined infrared power counting degrees
to the neutral Faddeev-Popov fields and the off-shell infrared
existence of higher-order Green functions is endangered (see \citere{ekhabil}).

The plan of the paper is as follows: In section 2 we give the classical
action of the electroweak SM including the gauge-fixing and ghost sector
in the on-shell parameterization. The gauge fixing is constructed in such
a way that it is compatible with the Ward identities of rigid symmetry and
the local abelian $U(1)$ Ward identity. The latter identity is crucial for
continuing the Gell-Mann Nishijima relation to higher orders.
In section 3 we give the Slavnov-Taylor identity 
in the tree approximation and show how the independent ghost mass ratio
can be consistently included.
In section 4 we derive the CS equation by constructing
the invariant differential operators. In section 5 we apply the
CS equation to different 1-loop vertices, calculating in this way
the $\beta$-functions and anomalous dimensions in one-loop order in
a general gauge. As an application at two-loop order, the leading
logarithms of the photon self-energy are investigated.
In section 6 we compare the results with the symmetric theory and
with QED of charged fermions. In section 7 we give our conclusions.
The appendix contains a list of free field propagators determined
in the general
linear $R_\xi$~gauge.

\newsection{The classical action of the electroweak Standard Model}

In order to set the general framework and to fix the notation we first give 
the classical action of the SM in the on-shell parameterization.
In our conventions we follow closely the ones used in \citere{Dehab}.

The Standard Model of electroweak interactions is a non-abelian gauge theory
with the non-semisimple gauge group
$SU(2) 
\times U(1) $.
It comprises four vector fields $V_{\mu , a}, a = +,-, Z,A $: the charged
bosons  $V_{\mu,{ \pm }} \equiv W_{\mu,\pm } $ with mass
$M_W$ and electric charge $ {\pm} 1 $, the neutral boson
$V_{\mu,Z}\equiv Z_\mu $ with mass $M_Z$ and the massless photon 
field $ 
V_{\mu,A} \equiv A_\mu $. The masses of the vector  bosons are generated
by spontaneous symmetry breaking via the Higgs mechanism.
The SM contains a complex scalar doublet 
\begin{equation}
 \Phi \equiv\left(
    \begin{array}{c}
      \phi^+(x)\\
    \frac  1{\sqrt 2}(H(x) + i\chi(x))
    \end{array}
  \right) 
\qquad \tilde \Phi \equiv i\tau_2 \Phi^* = \left(
    \begin{array}{c}
            \frac 1{\sqrt 2}(H(x) - i\chi(x)) \\
            -\phi^-(x) 
    \end{array}
  \right) ,
\end{equation}
where $H $ is the physical Higgs field with mass $m_H$, and
$ \phi^+, \phi^- $ and $\chi $ are the unphysical would-be 
Goldstone bosons.

In the fermion sector there are the left-handed fermion doublets,
the lepton and quark doublet
\begin{equation}
F^{L}_{\delta,i} \equiv F^{L}_{l,i} , F^{L}_{q,i} \qquad
F^{L}_{l,i} = \left( \begin{array}{c} \nu^L _i \\
                                    e^L _i \end{array} \right)
\qquad
F^{L}_{q,i} = \left( \begin{array}{c} u^L _i \\
                                    d^L _i \end{array} \right)
\end{equation}
and the right-handed singlets
\begin{equation}
f_i^R =  e_i^R, u_i^R , d_i^R .
\end{equation}
Here $i$ denotes the family index; $ \nu_i $ stands for neutrinos,
$e_i$ for charged leptons
 with mass $ m_{e_i}$ and electric charge $Q_e = -1$,
 $u_i$ and $ d_i$ for up and down-
type quarks with mass $ m_{u_i}$ and $m_{d_i}$ and electric charge $
Q_{u}= \frac 23$
and $Q_{d} = - \frac 13 $.
Since we are mainly interested in the CS functions of the vector sector
we do not consider mixing between different families, especially
we assume CP-invariance throughout the paper.

For convenience we give the classical action of the SM
as it arises after spontaneous breaking of the symmetry in terms
of the physical fields, i.e.~in mass and charge eigenstates.
The free parameters are the masses of the fields given above and one
coupling, which is chosen according to a QED-like parameterization:
\begin{equation}
\label{para}
M_W , M_Z , m_H , m_{f_i} , e .
\end{equation}
We introduce  the notation
\begin{equation}
\label{wein}
\cos {\theta_W} = \frac {M_W}{M_Z} ,
\end{equation}
which relates the weak mixing angle to the mass ratio of the $W$- and
$Z$-bosons. In higher orders the masses and also the field renormalizations
have to be fixed by appropriate normalization conditions for the
two-point functions. In a QED-like parameterization
the coupling 
can be fixed as the interaction strength of the photon to
the electromagnetic current in the Thompson limit, where it is 
determined by the fine structure constant.
\begin{equation}
\label{coupl}
\bar u (p)  \Gamma _{ee{A_\mu}}(p,p,0) u(p) \big| _{p^2 =  m_e^2} 
= i e \bar u(p) \gamma_\mu u(p) .
\end{equation}
The classical action can be decomposed into a gauge-invariant part
$\Gamma _{GSW} $ and the gauge-fixing and ghost part, which are constructed
to be BRS-invariant. The gauge-invariant part of the
action is given by:
\begin{equation}
\Gamma_{GSW} = \Gamma_{YM} + \Gamma_{scalar} + \Gamma_{ferm} \; ,
\end{equation}
\begin{eqnarray}
  \Gamma_{YM} & = &
  -\frac 1 4 \intd G_a^{\mu\nu}\tilde{I}_{aa'}G_{\mu\nu a'}\\
  \Gamma_{scalar}&=&\intd\Bigl( (D^\mu(\Phi +   {\hbox{v}}
 ))^\dagger D_{\mu}(\Phi
+   {\hbox{v}}) 
  -\frac 1 8\frac{m_H^2}{M^2_W} {e^2 \over \sws}
(\Phi^\dagger\Phi +   {\hbox{v}} 
 ^\dagger \Phi
+ \Phi^\dagger   {\hbox{v}})  ^2 \Bigr) \\  
\Gamma_{ferm} &=& \sum_{i=1}^{N_F} \intd \Bigl( \overline
 {F^L_{l,i}} i \Ds F^L_{l,i}
       + \overline {F^L_{q,i}} i \Ds F^L_{q,i} + \overline
 {f^R_i} i \Ds f^R_{i}\\ 
                  & & 
\phantom{\sum \intd} -  {e  \over M_W \sqrt 2 \sin \theta_W
} \bigl( m_{e_i} 
\overline
 {F^L_{l,i}} (\Phi +   {\hbox{v}}) e^R_i
                 +  m_{u_i}  \overline
 {F^L_{q,i}} (\Phi +   {\hbox{v}}) u^R_i \nonumber \\
                &  &
\phantom{\sum\intd}  \qquad + m_{d_i}  \overline
 {F^L_{q,i}} (\tilde \Phi + \tilde{  {\hbox{v}}}) d^R_i + \hbox{h.c.}
 \bigr) \Bigr) ,
\nonumber 
\end{eqnarray}
where $N_F$ is the number of fermion generations, and
$ { \hbox{v}} $ denotes the shift of the scalar field doublet, which 
generates the masses of the particles:
\begin{equation}
\label{shift}
{ \hbox{v}} = \left( \begin{array}{c} 0 \\
                                 \hbox{$\frac 1{\sqrt 2}$}
   v \end{array} \right)
\quad \hbox{with} \quad v = \frac 2e M_Z \cos\theta_W \sin \theta_W \; .
\end{equation}
It has a component into the direction of the physical Higgs field. Assigning
to the fields a definite transformation behavior under C,P and T
the action can be shown to be CP-invariant. (A table of quantum numbers
for all fields of the Standard Model can be found in \citere{ekhabil}.)

The field strength tensor and the covariant derivative have the form
\begin{eqnarray}
  G^{\mu\nu}_a & = & \partial ^\mu V_a^{\nu} - \partial^{\nu} V_a^\mu
  + \frac e{\sin{\theta_W}}
 \tilde{I}_{aa'} 
  {f}_{a'bc} V^{\mu}_b V^{\nu}_c\\
  D_{\mu}\Phi &=& \partial_{\mu}\Phi - i \frac e{\sin{\theta_W}}
 \frac{{\tau}_a(G_s)}{2} \Phi V_{\mu a}
\\
  D_{\mu} F^L_{\delta,i} &=& \partial_\mu F^L_{\delta,i} - i 
\frac e{\sin{\theta_W}}
 \frac{{\tau}_a(G_\delta)}{2} F^L_{\delta,i} V_{\mu a} \qquad \delta= l,q\\
  D_{\mu} f_i^R &=& \partial_\mu f_i^R  +
 i e Q_f \frac {\sin{\theta_W}}{\cos{\theta _W}}   f_i^R Z_\mu +
            ie Q_f f_i^R A_\mu \; .
\end{eqnarray}
We use the summation convention for the roman indices $a,b,c$
with values $+,-, Z,A$ and have introduced convenient notations.
The tensor
\begin{equation}
   {f}_{abc}  =  \left\{
    \begin{array} {ccc}
       {f}_{+-Z}&=&- i \cos\theta_W\\
       {f}_{+-A}& =&i \sin\theta_W 
    \end{array}\right.
\end{equation}
is completely antisymmetric and the matrices $ \tau_a\ \ (a=
+,-,Z,A)$ form a representation of $SU(2) \times U(1)$ according to
\begin{equation}
\label{taualg}
  \left[\frac{ {\tau}_a}2, \frac{ {\tau}_b}2\right]= 
  i  {f}_{abc} 
  \tilde{I}_{cc'} \frac{ \tau_{c'}}2 \; .
\end{equation}
They are explicitly given by $(\tau_i $, $i = 1,2,3$, are the Pauli
matrices)
\begin{eqnarray}
\label{taudef}
  {\tau}_+  
= \hbox{$\frac 1{\sqrt{2}}$} (\tau_1 + i\tau_2) &  &
  {\tau}_Z(G)  =\tau_3 \cos\theta_W + G {\mathbf 1} \sin\theta_W
    \nonumber\\
  {\tau}_-\!
 =\hbox{$\frac 1{\sqrt{2}}$} (\tau_1 - i\tau_2 ) &  &
  {\tau}_A (G) = - \tau_3 \sin\theta_W \! + G{\mathbf 1}\cos\theta_W \; .
\end{eqnarray}
These combinations depend on the abelian coupling  $G$, which is 
not determined by the algebra.
It is related to
the weak hypercharge $Y_W$
and accordingly to the electric charge $Q_f$ of the
particles:
\begin{equation}\label{hyper}
  G_k = - Y^{(k)}_W \frac{\sin\theta_W}{\cos\theta_W} \qquad
Y_W^{(k)}  =  \left\{
    \begin{array} {ccc}
       &1 & \hbox{for the scalar ($k=s$)} \\
       &$-1$& \hbox{for the lepton doublets ($k=l$)} \\
       & \hbox{$\frac 13$} & \hbox{for the quark doublets ($k=q$) . } 
    \end{array}\right. 
\end{equation}
The matrix $\tilde{I}_{aa'}$ guarantees the charge neutrality of the
classical action
\begin{eqnarray}
\label{tildeI}
\tilde{I}_{+-}&=&\tilde{I}_{-+}=\tilde{I}_{ZZ}=\tilde{I}_{AA}=1\\
\tilde{I}_{ab}& =&0 \mathrm{\ else} . \nonumber
\end{eqnarray}

The action $\Gamma_{GSW}$ is manifestly invariant under
$SU(2) \times U(1)$ gauge transformations,
if one includes the shift of the Higgs field
into the transformation. The transformation behavior of the physical
fields can be read off from the covariant derivatives.

In order to quantize the SM the gauge is fixed 
in such a way that renormalizability is guaranteed by power
counting. 
Taking the usual linear  $R_\xi$~gauges we choose the following
gauge-fixing functions, which are the most general ones having
definite transformation with respect to CP:
\begin{eqnarray}
\label{Fgen}
F_{\pm}&\equiv&\partial_{\mu}W^{\mu}_{\pm} \mp iM_{W}\zeta_{W}\phi^{\pm}
\nonumber{}\\
F_Z&\equiv& \partial_{\mu}Z^{\mu}-M_{Z}\zeta_{Z}{{\chi}}\\
F_A&\equiv & \partial_{\mu}A^{\mu} - M_Z \zeta_{A}{{\chi}} . \nonumber 
\end{eqnarray}
The mass terms of the would-be Goldstone fields are introduced in order
to remove non-integrable infrared divergencies from the propagators.
Coupling the gauge-fixing functions 
to a Lagrange multiplier field $B_a, a= +,-,Z,A$, with dimension 2 
and odd under CP-transformations, the gauge fixing reads
\begin{eqnarray}
\Gamma_{ {g.f.}}&= 
& \intd\left(\frac 12 \xi _{ab} B_a B_{b}+ B_a \tilde{I}_{ab} F_{b} 
\right) .
\label{eq:gf}
\end{eqnarray}
It can be transformed into its usual form by eliminating the
$B_a$-fields via their equations of motion:
\begin{equation}
\label{Rxi}
\frac{\delta \Gamma}{\delta B _a } 
= \xi _{ab} B_b + \tilde I _{ab} F_b \, {\buildrel \ast \over =} \, 0 .
\end{equation}
The gauge fixing breaks  gauge invariance and also its
integrated version, the rigid $SU(2) \times U(1)$ symmetry, which
is obtained by taking the infinitesimal transformation parameters
of gauge transformations as constants.
Therefore the unphysical fields, the longitudinal parts of the vectors
and the would-be Goldstones,  interact with the physical fields
violating thereby unitarity.
For this reason one has to introduce the Faddeev-Popov  fields
$c_a , a= +,-, Z,A$, with ghost charge 1 and the respective 
antighosts $\bar c_a , a= +,-, Z,A $, with ghost charge $-1$ and
has to add the ghost part in such a way that the classical
action is invariant under BRS transformations:
\begin{eqnarray}
\label{brs}
{\mathrm s} V_{\mu a}&=&\partial _{\mu}c_a + 
\frac e{\sin \theta _W} \tilde I _{aa'} f_{a'bc}V_{\mu
  b}c_c\nonumber{}\\
{\mathrm s} c_{a}&=&- \frac e{2 \sin \theta_W } \tilde I _{aa'} 
 f_{a'bc}c_bc_c\nonumber{}\\
{\mathrm s} \Phi &=& i \frac e{\sin{\theta_W}}
\frac{{\tau}_a(G_s)}{2} (\Phi+ {\mathrm v}) c_{
 a} \nonumber  \\
{\mathrm s} F^L_\delta & =& i
\frac e{\sin{\theta_W}}
 \frac{{\tau}_a(G_\delta)}{2} F^L_\delta c_{ a} \qquad \delta = l,q\\
{\mathrm s}  f^R&= & - i e Q_f \frac {\sin{\theta_W}}{\cos{\theta _W}}   f^R
  c_Z-            ie Q_f f^R c_A \nonumber \\
{\mathrm s} \bar c_a &=& B_a \nonumber \\
{\mathrm s} B_a &=& 0 . \nonumber
\end{eqnarray} 
Having formulated the gauge fixing with the auxiliary fields
$B_a$, the BRS transformations are nilpotent on all fields
\begin{equation}
{\mathrm s}^2 = 0 .
\end{equation}
Requiring the classical action to be BRS-invariant
\begin{equation}
\label{gacl}
{\mathrm s} \Gamma_{cl} = 0 \quad \hbox{with} \quad
 \Gamma_{cl} = \Gamma_{GSW} + \Gamma_{g.f.} + \Gamma_{ghost} ,
\end{equation}
 the ghost action is determined 
\begin{equation}
\Gamma_{{ghost}} = \intd \left(- \bar c_a \tilde I _{ab} {\mathrm s} F_b 
\right) .
\end{equation}
The bilinear terms are given explicitly by
\begin{eqnarray}
\label{ghostbil}
\Gamma^{(bil)} _{ghost} &=& \intd \Bigl(- \bar c _a \Box \tilde I_{ab} c_b - 
  \zeta _W M_W^2 (\bar c_+ c_- + \bar c_- c_+)  \\ 
& &\phantom{\intd} -\zeta _Z M_Z^2 \bar c_Z c_Z  -
\zeta _A M_Z^2 \bar c_A c_Z  \Bigr) .\nonumber
\end{eqnarray}  
With the help of
BRS invariance one is able to prove unitarity of the physical
S-matrix in the tree approximation.
 It is therefore the relevant symmetry for 
quantization and renormalization because it fixes the interactions
amongst the unphysical fields in such a way that the complete 
action is renormalizable and eventually the physical S-matrix unitary
 \cite{BRS75, KUOJ78}.

The gauge-fixing parameters are not specified by BRS invariance.
In general $\xi_{ab}$ is an arbitrary symmetric matrix and
$\zeta_a$  are arbitrary parameters.
They have to be restricted by normalization conditions on
the ghost propagators or additional symmetries.
A natural choice in the tree approximation is 
\begin{equation}
\label{fixrig}
\xi_{ab} = \xi \tilde I_{ab} \qquad \zeta_W = \zeta_Z \qquad \zeta_A =0,
\end{equation}
which makes the  propagators of the longitudinal vectors and 
of the Faddeev-Popov ghosts diagonal.
 If we constrain the gauge fixing according to
(\ref{fixrig}), rigid invariance is still broken
 by the mass terms and, moreover, the breaking transforms covariantly
in such a way that it can be controlled by introducing 
an external scalar doublet
\begin{equation}
\hat \Phi = {\hat \phi^ + \choose \frac 1 {\sqrt 2 }
( \hat H + i \hat \chi )}
\end{equation} 
with the same quantum numbers as the scalar doublet \( \Phi \), but 
which is BRS-transformed into an external doublet ${\mathrm q}$ with ghost 
charge 1 (cf.~\citeres{ekhabil,KS})
\begin{equation}
{\mathrm s} \hat \Phi = {\mathrm q} \qquad 
{\mathrm s} {\mathrm q} = 0 .
\end{equation}
In the 
most general linear gauge fixing invariant with respect
to rigid symmetry transformations 
the gauge parameters are restricted as follows
\begin{eqnarray}
\label{xidef}
\xi_{ab} &=& \xi \tilde I_{ab} + \hat\xi 
\left(
\begin{array}{cccc}
      0& 0& 0& 0 \\
      0& 0& 0& 0 \\
      0& 0& \sws& \sw \cw\\
      0& 0& \sw \cw& \cws
    \end{array}
\right) \\
\label{zetadef}
\zeta_W & = & \zeta  \\
\zeta_Z & =  &\zeta \cos \theta_W ( \cos \theta_W - \hat G
\sin \theta_W ) \label{zetadef2} \\
\zeta_A & = & - \zeta\cos \theta_W (\sin \theta _W - \hat G \cos\theta_W) . 
\label{zetadef3}
\end{eqnarray}
Then the gauge fixing \refeq{eq:gf} reads explicitly (including also
the external scalars)
\begin{eqnarray}
\label{fixext}
\Gamma_{{g.f.}}&\hspace{-3mm}=&\hspace{-3mm}\intd 
\biggl( \frac 12\xi B_a \tilde I _{ab} B_b +
 \frac 12 \hat \xi (\sw B_Z + \cw B_A )^2 +
B_a \tilde I_{ab} \partial V_b \nonumber\\
&\hspace{-3mm} &\hspace{-3mm}-\frac {ie}{\sin \theta _W }
\bigl( (\hat \Phi  +  \zeta {\mathrm v}
)^\dagger \frac { \tau _a (\hat G) } 2  B_a ( \Phi +{\mathrm v})
- (\Phi + {\mathrm v})^\dagger \frac {\tau _a (\hat G) } 
2  B_a  (\hat \Phi + \zeta 
{\mathrm v})\bigr) \biggr) ,
\end{eqnarray} 
with $\mathrm v $ given in (\ref{shift})
and $\tau _a(G) $  in (\ref{taudef}). 
Concerning the external scalar doublet $\hat \Phi$ and the quantum scalar
doublet $\Phi$ 
this gauge-fixing term is equivalent to the one used in
the background-field method~\cite{bfmgf,bfmlong}.

It is seen that the would-be Goldstone fields $\phi^\pm $ and $\chi$ as
well as the massive Faddeev-Popov ghosts get their masses via the 
shift of the external scalar fields $\zeta \mathrm v$. 
The gauge parameter $\hat \xi$ as well as 
the abelian coupling $\hat G$ are not determined by rigid symmetry,
gauge invariance or BRS symmetry. The parameterization chosen in
(\ref{fixrig}) reads now:
\begin{equation}
\label{gmin}
\hat \xi = 0 \quad\hbox{and} \qquad
\hat G = - \frac {\sin \theta_W} {\cos \theta _W } . 
\end{equation}
It turns out, however, that the minimal choice 
(\ref{gmin})
is not stable
under renormalization as  will be seen from
the Callan-Symanzik equation. In particular, the ghost mass ratio
is independently renormalized from the vector mass ratio.
In view of the investigation of higher-order contributions it is therefore
important to work in the general gauge specified in \refeq{fixext}.

\newsection{Quantization}

For a systematic treatment of quantization and renormalization
one expresses invariance under BRS transformations
(\ref{brs}) and rigid symmetry 
in the form of functional
operators, the Slavnov-Taylor identity and the
Ward identities of rigid symmetry. 

Since the BRS transformations include non-linear field transformations
in propagating fields, they have to be coupled to external fields.
In order to avoid double definitions of the insertions 
$ c_+ c_-  $ and $ V_+ c_- - V_ - c_+$ one has
to split off the linear $U(1)$-transformations explicitly \cite{BABE78,
 ekhabil}.
We introduce the external field action in the following form:
\begin{eqnarray}
\label{gaext}
\Gamma_{ext.f.} & = & \intd \Bigl( \rho ^\mu _+ {\brs} W_{\mu,-} +
                           \rho ^\mu _- {\brs} W_{\mu,+ } +
                           \rho ^\mu _3  (\cw {\brs} Z_{\mu} - \sw 
{\brs} A_\mu) \\
               &   & \! \phantom{\intd} 
+ \sigma _+ {\brs} c_{-} +
                                        \sigma _- {\brs} c_{+} +
                               \sigma _3 (\cw {\brs}  c_Z - \sw {\brs} c_A) 
\nonumber \\
                              &   & \! \phantom{\intd} 
       + Y^\dagger {\brs} \Phi  + ({\brs} \Phi) ^\dagger Y \nonumber \\
    & & \! \phantom{\intd}
       + \sum _{i=1}^{N_F} \bigl(\sum_{\delta = l,q}
       \overline {\Psi ^R _{\delta,i}} {\brs} F^L _{\delta,i}  + 
       \sum_{f}       \overline {\psi ^L _{f,i}}
{\brs}  f^R_i + \hbox{h.c.} \bigr)\Bigr) .
\nonumber
\end{eqnarray}
The external fields $\rho^\mu_\alpha$ and $\sigma_\alpha, \alpha = +,-,3$,
are  $SU(2)$-triplets with ghost charge $-1$ and $-2$, respectively.
The external field $Y$ is a complex scalar doublet 
with ghost charge $-1$, $\psi_{f,i} ^L $ denotes  external left-handed 
spinor singlets with ghost charge $-1$,
\begin{equation}
\psi^L_{f,i} \equiv \psi^L _{e,i} , \; \psi^L _{u,i} , \; \psi^L _{d,i} ,
\end{equation}
whereas $\Psi_{\delta,i} ^R $ denotes  external right-handed spinor doublets
\begin{equation}
\Psi^R _{\delta, i} \equiv \Psi^R _{l,i} ,\, \Psi^R _{q,i} \qquad
\Psi^R _{l,i}  =
 \left( \begin{array}{c} \psi ^R _{\nu,i} \\
                                \psi^R _{e,i} \end{array} \right)
\qquad
\Psi^R _{q,i}  =
 \left( \begin{array}{c} \psi ^R _{u,i} \\
                                \psi^R _{d,i} \end{array} \right).
\end{equation}
The transformation under discrete symmetries is assigned in such a way
that the external field part is  neutral and CP-invariant.

The classical action corresponds to the lowest order of the  
perturbative expansion of 1PI Green functions and 
one can read off the respective functional operators of the defining
symmetries in the tree approximation. 
Including the external field part (\ref{gaext}) into the classical action 
(\ref{gacl})
one is able to encode the  BRS transformations (\ref{brs}) 
in the Slavnov-Taylor (ST) operator:
\begin{eqnarray}
\label{ST}
{\cal S}
(\Gamma ) &=& \intd \biggl(
\bigl(\sw \partial _\mu c _Z + \cw \partial_\mu c_A\bigr)
             \Bigl(\sw {\delta \Gamma \over \delta Z_\mu } + \cw
       {\delta \Gamma \over \delta A_\mu} \Bigr) \\  
 & & +     {\delta \Gamma \over \delta \rho^\mu_3 }
              \Bigl(\cw {\delta \Gamma \over \delta Z_\mu } - \sw
       {\delta \Gamma \over \delta A_\mu} \Bigr) 
       + {\delta \Gamma \over \delta \sigma _3 }
              \Bigl(\cw {\delta \Gamma \over \delta c_Z } - \sw
       {\delta \Gamma\over \delta c_A} \Bigr) \nonumber \\ 
& &      + {\delta \Gamma \over \delta  \rho^\mu _+ }
               {\delta \Gamma \over \delta W_{\mu -} }
      + {\delta \Gamma \over \delta \rho^\mu _- }
               {\delta \Gamma \over \delta W_{\mu + } }
+  {\delta \Gamma \over \delta \sigma _+ }
               {\delta \Gamma \over \delta c_{-} }
+  {\delta \Gamma \over \delta \sigma _- }
               {\delta \Gamma \over \delta c_{+} } +
{\delta \Gamma \over \delta Y^\dagger}{ \delta \Gamma \over \delta \Phi } +
{\delta \Gamma \over \delta \Phi^\dagger}{ \delta \Gamma \over \delta Y } 
 \nonumber \\
&& + \sum_{i=1}^{N_F} \Bigl({\delta \Gamma \over \delta 
\overline{\psi^L_{f,i}}}
{ \delta \Gamma \over \delta f^R_i }
+ {\delta \Gamma \over \delta \overline{\Psi^R_{\delta,i}} }
{ \delta \Gamma  \over \delta F^L_{\delta,i }} + 
  \hbox{h.c.} \Bigr)  \nonumber \\
& & + B_a {\delta \Gamma \over \delta \bar c_a }  
   + {\mathrm q}{\delta \Gamma \over \delta \hat \Phi  }  + 
{\delta \Gamma \over \delta \hat \Phi ^\dagger } {\mathrm q}^\dagger \biggr) .
   \nonumber
\end{eqnarray}
The ST identity of the tree approximation 
\begin{equation}
{\cal S} (\Gamma_{cl}) = 0 
\end{equation}
is fulfilled by construction. 

Rigid symmetry can be formulated in terms of  linear, integrated 
Ward operators which satisfy the $SU(2) \times U(1) $ algebra
\begin{eqnarray}
\label{alg}
\bigl[ {\cal W }_\alpha , {\cal W } _\beta \bigr] & =& \varepsilon 
_{\alpha\beta\gamma} \tilde I_{\gamma \gamma'} {\cal W} _{\gamma'} \\
\bigl[ {\cal W }_\alpha , {\cal W } _4 \bigr] &=& 0 . \nonumber
\end{eqnarray}
The Greek indices are $SU(2)$-group indices and
run over $+,-, 3$, the Ward operator ${\cal W}_4$ corresponds to the
transformation under  the $U(1)$-group and commutes therefore with all
 Ward operators. The tensor $\varepsilon_{\alpha \beta \gamma} $ denotes
the structure constants of charged $SU(2)$ and is completely antisymmetric:
\begin{equation}
\label{scon}
       {\varepsilon}_{+-3} =  - i .
\end{equation}
 Vector fields,  
 Faddeev-Popov ghosts and the auxiliary fields $B_a$ transform
according to the adjoint representation, whereas all the scalars
transform according to the fundamental representation.
 The external fields $\rho^\mu_\alpha $ and $\sigma_\alpha, \alpha = +,-,3$,
are only transformed under $SU(2)$.
We thus arrive at
\begin{equation}
\label{wi}
{\cal W} _\alpha \Gamma _{cl} = 0 
\quad \hbox{and} \quad {\cal W} _4 \Gamma _{cl} = 0 ,
\end{equation}
where the Ward operators of the tree approximation are given by
\begin{eqnarray}
\label{wardna} 
{\cal W}_\alpha & = & \tilde I_{\alpha\alpha'} 
\intd 
 \biggl( \Bigl(  V^\mu _b \hat\varepsilon _{bc,\alpha'}  \tilde I _{cc'}
\frac{\delta}{\delta V^\mu _{c'}} 
 + \{  c, B, \bar c \} \Bigr) \\
& & \phantom{ \tilde I_{\alpha\alpha'} }
+ \Bigl(
 \rho^\mu _\beta \varepsilon _{\beta\gamma \alpha'}  \tilde I _{\gamma \gamma'}
\frac{\delta}{\delta \rho^\mu _{\gamma'}} + \{\sigma \}\Bigr)  \nonumber \\
&  &  \phantom{ \tilde I_{\alpha\alpha'} } + \Bigl(
  i (\Phi + {\mathrm v}) ^\dagger
\frac { \tau _{\alpha'} } 2  \frac{\overrightarrow 
\delta}{\delta \Phi^\dagger
}
 -  i \frac{\overleftarrow
\delta}{\delta \Phi} \frac { \tau _{\alpha'} } 2  (\Phi +{\mathrm v})  
+  \{ Y , \hat \Phi + \zeta {\mathrm v}, q \} \Bigr)\nonumber \\
& &  \phantom{ \tilde I_{\alpha\alpha'} }
 + \sum_{\delta, i} \Bigl(
 i \overline{F ^L_{\delta,i}} \frac {\tau _{\alpha'} } 2 
 \frac{\overrightarrow 
\delta}{\delta \overline{ F^L_{\delta,i} }}
 -  i \frac{\overleftarrow
\delta}{\delta F^L_{\delta,i}} \frac { \tau _{\alpha'} } 2  F^L_{\delta,i}  
+  \{ \Psi^R _{\delta,i} \} \Bigr) \biggr)\,  . \nonumber
\end{eqnarray}
Fields in curly brackets in \refeq{wardna} denote that these fields are 
transformed in the same way as the one explicitly given in the respective
line of the formula.
 The matrices $\tau_\alpha = \tau _+ ,\tau _- ,\tau _3 $ are 
the Pauli matrices of the charged representation of $SU(2)$ (\ref{scon}).
The tensor $\hat \varepsilon_{bc,\alpha}$, which
governs the transformation of the vector fields,
is antisymmetric in the first two
indices. These indices are field indices and are generated by 
rotating   the neutral $SU(2)$-fields and
the abelian fields by the weak mixing angle $\theta_W$ into on-shell fields:
\begin{equation}
   {\hat \varepsilon}_{bc,\alpha}  =  \left\{
    \begin{array} {ccc}
       {\hat \varepsilon}_{Z+,-}&=&- i \cos\theta_W\\
       {\hat \varepsilon}_{A+,-}& =&i \sin\theta_W \\
       \hat \varepsilon_{+-,3}& =& -i 
    \end{array}\right. .
\end{equation}

In the electroweak Standard Model there are several rigid abelian operators
$\cW _4$
which commute with the $SU(2)$ operators (\ref{wardna}). 
The respective symmetries of the classical action
 correspond to the conservation of electromagnetic charge
($  \cW_{em} - \cW_3 $) 
 and conservation of  lepton  and quark family number ($\cW_{l_i}$ and 
$\cW_{q_i}$), 
\begin{equation}
{\cal W}_4 \Ga_{cl} = 0 \quad \mbox{where} \quad
{\cal W}_4 \equiv \bigl({\cal W}_{em} - {\cal W}_3  \bigr) + \sum_{i}^ {N_F}
g_{l_i} {\cal W}_{l_i} + g_{q_i} {\cal W}_{q_i} . 
\end{equation}
Here ${\cal W}_{em}$ is the usual electromagnetic charge operator; explicit
expressions are given in \citere{ekhabil}.
From the corresponding classically conserved currents only the electromagnetic
current is gauged. 
 When the gauge-fixing sector and ghost sector is constructed
in accordance with rigid symmetry (\ref{fixext}), then it is possible
 to establish 
 the local abelian Ward identity  corresponding to electromagnetic
current conservation
\begin{equation}
\label{wardloc}
{\mathbf w}^Q_4 \Gamma_{cl}- \frac 1e 
\cw \Bigl(\sw \partial {\delta \Gamma_{cl}\over \delta Z} + \cw \partial 
{\delta \Gamma _{cl}\over \delta
A}\Bigr) = 
\frac 1e \cw (\sw \Box B_Z + \cw \Box B_A ) \, ,
\end{equation}
with
\begin{equation}
\label{GMN}
{\mathbf w}_4^Q 
= {\mathbf w}_{em} -  {\mathbf w}_3\, .
\end{equation}
The local operators ${\mathbf w}_3$ and ${\mathbf w}_{em}$
 are defined by taking away the integration from the 
rigid operators:
\begin{equation}
{\cal W}_3 = \intd {\mathbf w}_3 
\quad \hbox{and} \quad {\cal W}_{em} = \intd {\mathbf w}_{em} \; .
\end{equation}
The local Ward identity 
together with (\ref{GMN}) is the functional generalization of the
Gell-Mann Nishijima relation and allows to determine the weak
hypercharges of fermion dou\-blets and the electromagnetic
charges of fermion singlets. In higher orders this local Ward identity
plays an important role for a scheme-independent
definition of the abelian charges 
(for details see \citere{ekhabil}).\footnote{In \cite{GR98} 
the abelian charges are fixed
 by postulating an antighost 
equation. From  there   the local Ward identity  is defined by using the
consistency with the ST identity.}

In the procedure of renormalization and
quantization one has to construct the Green functions
in such a way that they satisfy simultaneously the
Slavnov-Taylor identity, the Ward identities of rigid symmetry
specified by the commutation relation (\ref{alg}), and the abelian local
Ward identity (\ref{wardloc}).
This problem has to be taken seriously
 due to the fact that the photon and the
respective Faddeev-Popov ghosts are massless and these massless particles
have non-abelian interactions. In order to ensure that
all integrals are infrared convergent for non-exceptional momenta
 in higher orders of perturbation theory
one has to supplement the usual normalization conditions, which
fix the free parameters (\ref{para}) and the wave function normalization,
 by the requirements
 that also the mixed 2-point functions of massive
and massless particles vanish at $p^2 = 0$:
\begin{eqnarray}
\label{infra}
\Gamma_{ZA}(p^2=0) = \Gamma_{AA}(p^2 = 0) &=& 0 \\
\Gamma_{\bar c _A c _Z}(p^2=0) =
\Gamma_{\bar c _Z c _A}(p^2=0) = \Gamma_{\bar c_ Ac _A}(p^2 = 0) &=& 0 .
\nonumber
\end{eqnarray}
A careful analysis of higher orders shows that on-shell conditions
which include (\ref{infra}) as well as corresponding conditions
at the mass of the massive fields
can be only fulfilled in agreement with the Slavnov-Taylor identity and the
Ward identities if one takes into account higher-order corrections 
 of the ST operator and the Ward operators~\cite{ekhabil}.
The $\beta$-functions
and anomalous dimensions 
of the 1-loop CS equation
are independent of these higher-order corrections.
 But one cannot completely stick to the tree
approximation  as defined by the parameterization (\ref{gmin})
and the BRS transformations (\ref{brs}) since the ghost mass ratio
turns out to be an independent parameter
of the Standard Model.
This means that we have to keep the
parameters of the gauge-fixing functions (\ref{fixext}) 
$\xi , \hat \xi , \zeta $ and
$\hat G$  as independent parameters. Keeping
$\hat G$ arbitrarily and using at the same time
the BRS transformation (\ref{brs}) brings about that the non-diagonal
 mass term $\bar c_A c_Z $ arises in the action (see~(\ref{ghostmass})).
Such  a non-diagonal mass term leads in higher orders to off-shell 
infrared divergent
contributions and prevents the introduction of
definite infrared degrees of
power counting for the ghosts. In order to remedy the situation one
has to introduce a ghost angle into the BRS transformations already
in the tree approximation,
which allows consistently to remove the infrared divergent 
contributions $\bar c_A c_Z$ from the classical action for
arbitrary ghost mass ratio and to derive the CS equation.

For the purpose of this paper we want to outline the definition
of the ghost angle in the tree approximation, whereas a detailed
analysis  especially of higher orders is given in \citere{ekhabil}.
  The linear $R_\xi $ gauge-fixing functions  (\ref{Fgen}) are
 restricted by 
rigid symmetry as given in (\ref{xidef})--(\ref{zetadef3}). They include
only four free parameters:
\begin{equation}
\label{gaugepar}
\xi, \, \hat\xi , \,  \zeta ,\,  \hat G .
\end{equation}
The parameters $\zeta$ and $\hat G$ are connected with  the ghost masses:
\begin{equation}
\label{ghostmass}
 -  \zeta 
M_W \! \! \intd \!\! \Bigl(  M_W (\bar c_+ c_- + \bar c_ - c_+ ) +
         M_Z  \bar c_Z c_Z  (\cos \theta_W
 - \hat G 
{\sin \theta_W } ) - M_Z \bar c_A c_Z (\sin \theta_W
+\hat G
{\cos \theta_W}) \Bigr) .
\end{equation}
Introducing in analogy to (\ref{hyper}) the notation
\begin{equation}
\hat G = -{\sg \over \cg} ,
\end{equation}
the ratio of the ghost masses is determined by
\begin{equation}
\label{gmr}
{\zeta_W M_W ^2\over \zeta_Z M_Z^2 } = {\cw \cg \over \cwg} ,
\end{equation}
where $\zeta_W M_W ^2$ 
is the mass  of the charged ghosts and
$\zeta_Z M_Z^2 $
the mass of the massive neutral ghosts. The ghost angle $\theta_G$
is uniquely determined for arbitrary ghost masses.
In order to be able to remove the non-diagonal ghost mass term for
arbitrary ghost mass ratio from the action one has to redefine the
neutral 
ghosts $c_Z$ and $ c_A $ as well as the antighosts 
$\bar c_Z$ and $\bar c_A$ by a non-diagonal matrix $\hat g_{ab}$:
\begin{equation}
\label{gred}
c_a \longrightarrow \hat g_{ab} c_b, \qquad  \bar c_a \longrightarrow
\bar c_b \hat g_{ba}^{-1} \; .
\end{equation}
This procedure is 
analogous
to the one which has to be carried out if one
constructs mass eigenstates in the vector sector
by introducing  the weak mixing angle. 
The matrix $\hat g_{ab}$ is determined from the normalization
conditions (\ref{infra}) up to two constants which we have fixed
for convenience:
\begin{eqnarray}
\hat g _{+-}  = 1 &\quad & \hat g _{-+} = 1 \nonumber\\
\hat g _{ZZ}  =  \cwg &\quad &\hat g_{AZ} = -\swg \\
\hat g_{ZA}  =  0 &\quad& \hat g_{AA} = 1 . \nonumber
\end{eqnarray}
In matrix notation it reads:
\begin{equation}
\hat g _{ab} = \left(
    \begin{array}{cccc}
      1& 0& 0& 0 \\
      0& 1& 0& 0 \\
      0& 0& \cwg & 0\\
      0& 0& - \swg& 1 
    \end{array}
  \right) .
\end{equation}
With these redefinitions the bilinear part of the Faddeev-Popov ghost action 
is diagonal also in the general gauge (\ref{fixext}): 
\begin{equation}
\label{ghostdiag}
\Gamma^{(bil)} _{ghost} = \intd \Bigl(- \bar c _a \Box \tilde I_{ab} c_b - 
  \zeta _W M_W^2 (\bar c_+ c_- + \bar c_- c_+)  -\zeta _Z M_Z^2 \bar c_Z c_Z 
\Bigr) .
\end{equation}  
Consequently the ghost propagators are diagonal and allow to assign a
well-defined  infrared degree of power counting to $Z$ and $A$-ghosts.
For arbitrary masses, however, the ghost angle enters the BRS transformations
and the ST identity and eventually also the ghost--vector
 interactions. 
Explicitly the ST operator reads
\begin{eqnarray}
\label{STG}
{\cal S}
(\Gamma ) &=& \intd \biggl(
\bigl(\sg \partial _\mu c _Z + \cw \partial_\mu c_A\bigr)
             \Bigl(\sw {\delta \Gamma \over \delta Z_\mu } + \cw
       {\delta \Gamma \over \delta A_\mu} \Bigr) \\  
 & & +     {\delta \Gamma \over \delta \rho^\mu_3 }
              \Bigl(\cw {\delta \Gamma \over \delta Z_\mu } - \sw
       {\delta \Gamma \over \delta A_\mu} \Bigr) 
       + {\delta \Gamma \over \delta \sigma _3 } 
              \Bigl(\cw {\delta \Gamma \over \delta c_Z } - \sg
       {\delta \Gamma\over \delta c_A} \Bigr) 
  \frac 1{\cwg}\nonumber \\ 
& & 
+ \Bigl(\cwg B_Z -\swg B_A\Bigr)
{\delta \Gamma \over \delta \bar c_Z } +
B_A{\delta \Gamma \over \delta \bar c_A }   \nonumber \\
& &      + {\delta \Gamma \over \delta  \rho^\mu _+ }
               {\delta \Gamma \over \delta W_{\mu,- } }
      + {\delta \Gamma \over \delta \rho^\mu _- }
               {\delta \Gamma\over \delta W_{\mu,+ } }
+  {\delta \Gamma \over \delta \sigma _+ }
               {\delta \Gamma \over \delta c_{-} }
+  {\delta \Gamma \over \delta \sigma _- }
               {\delta \Gamma \over \delta c_{+} } 
+ B_+{\delta \Gamma \over \delta \bar c_+ } +  
B_-{\delta \Gamma \over \delta \bar c_- }    
\nonumber \\
&& + \sum_{i=1}^{N_F} \Bigl(\sum_{f}
{\delta \Gamma \over \delta \overline{\psi^L_{f,i}}
}{ \delta \Gamma \over \delta f^R_i }
+ \sum _{\delta= l,q}{\delta \Gamma \over \delta \overline{\Psi^R_{\delta,i}}}
{ \delta \Gamma \over \delta F^L_{\delta,i} } + 
  \hbox{h.c.} \Bigr) +
\Bigl({\delta \Gamma \over \delta Y^\dagger}
{ \delta \Gamma \over \delta \Phi } 
+ {\mathrm q}
{\delta \Gamma \over S\delta \hat \Phi  }  + \hbox{h.c.} \Bigr)\biggr)
   \nonumber 
   \end{eqnarray}
and the ST identity is fulfilled in the tree approximation for arbitrary
ghost angle $\theta_G$:
\begin{equation}
{\cal S} (\Ga_{cl}) = 0
\end{equation}
Establishing the ST identity in higher orders, $\theta_W$ and $\theta_G$
get independent higher-order corrections in the on-shell scheme.
Their correct treatment is a
 necessary prerequisite
for obtaining
 infrared convergent higher-order
corrections for off-shell Green functions~\cite{ekhabil}.

The ghost angle enters also the Ward identities of
rigid $SU(2)$-transformations (\ref{wi}),(\ref{wardna})
 via the redefinitions 
(\ref{gred}). It is worth to note that the algebra
(\ref{alg}) remains unchanged
by such field redefinitions.
 
The usual formulation of the gauge-fixing sector without the $B_a$
fields is achieved if one eliminates the $B_a$~fields via
their equations of motion from the gauge-fixing action
(\ref{fixext}) and inserts the result 
into the gauge fixing as well as in the ST identity (\ref{STG}).
The above discussion concerning the ghost angle is independent from
the formulation of the gauge fixing with or without $B_a$~fields.


\newsection{The Callan-Symanzik equation}

The Callan-Symanzik equation describes the response of the Green
functions to the scaling of all momenta by an infinitesimal factor.
The dilatational operator, which is just the scaling operator, 
acts on the 1PI Green functions in the same way as the differentiation
with respect to all the mass parameters of the theory:
\begin{eqnarray}
& &{\cal W}^{D} \Gamma = - m\partial _m \Gamma \quad\hbox{with} \\
& & m\partial_m \equiv {M_W}\partial_{M_W} +
{M_Z}\partial_{M_Z} +
{m_H}\partial_{m_H} + \sum_{i=1}^{N_F}\sum_f
{m_{f_{i}}}\partial_{m_{f_i}} +\kappa \partial_\kappa \; . \nonumber
\end{eqnarray}
Here
$\kappa$ is a normalization point which is introduced in order to
fix the on-shell infrared divergent residua of charged particles off-shell
without introducing a photon mass term. 

In the SM dilatations are already broken in the tree 
approximation by the mass terms of the fields and the 3-dimensional
interactions.
 Due to the spontaneous symmetry breaking
all the masses of the physical fields are generated by the shift
of the Higgs field. According to the construction of the 
gauge-fixing sector using rigid symmetry (\ref{fixext}),
the ghost masses are generated by the shift of 
the external Higgs field. In the tree approximation one gets therefore
the expression
\begin{equation}
\label{cstree}
m\partial_m \Gamma_{cl} = \intd v \Bigl( {\delta \Gamma_{cl}\over \delta H } +
\zeta {\delta \Gamma_{cl}\over \delta \hat H } \Bigr) +  
\frac {m_H^2}2  \Delta_{inv} 
\equiv\intd v \Bigl( {\delta \over \delta H } +
\zeta {\delta \over \delta \hat H } + \alpha_{inv}
{\delta \over \delta \hat \varphi _o} \Bigr) \, \Gamma_{cl} .
\end{equation}
Here
$\Delta_{inv}$ is the 2-dimensional BRS- and rigid-invariant 
scalar polynomial
\begin{equation}
\Delta_{inv} \equiv \intd( 2 \phi^+ \phi^- + \chi ^2 + H^2 + 2 v H) 
= \intd(\Phi^ \dagger \Phi + {\mathrm v}^\dagger \Phi + \Phi^\dagger
{\mathrm v}  ) \, ,
\end{equation}
which we couple to an external invariant scalar $\hat \varphi_o$.
For proceeding to higher orders it is important to note 
that the differential
operators $m\partial _m $ as well as $ {\delta \over \delta H } $
and $
 {\delta \over \delta \hat H }$ are BRS-symmetric operators
 and
 have  a certain covariance with respect
to rigid symmetry:
\begin{eqnarray}
\label{wsymsoft}
&&\Bigl[{\cal W }_\a ,m \partial _m\Bigr] = 
\Bigl[{\cal W }_\a,\intd v\Bigl( {\delta \over \delta H } +
\zeta {\delta \over \delta \hat H } \Bigr) \Bigr], \qquad \a= +,-,3,4.
\end{eqnarray}
The soft breaking is completely characterized by these symmetries and
takes therefore the same form to all orders of perturbation theory,
if one takes higher-order corrections of the shift and the coefficient
$\alpha_{inv}$ into account:
\begin{equation}
v = \frac 2 e M_Z \cw \sw + O(\hbar), \qquad 
\alpha_{inv} = \frac {m^2_H} {2v} + O(\hbar) .
\end{equation}

In higher orders the dilatations are not only broken by the
soft mass terms but also by hard terms, the dilatational anomalies.
The importance of the Callan-Symanzik equation is founded in the fact that
these anomalies can be 
absorbed into differential operators with respect to fields
and with respect to the independent parameters of the model.
Their coefficients are  the anomalous 
dimensions and the $\beta$-functions. 
In this way the CS equation determines the parameters that
are independently renormalized in  a scheme-independent way. 

The dilatational
anomalies at one-loop order are normalization-point-independent,
but the differential operators introduced depend on the 
parameterization and the specific form of the breaking mechanism.
They are essentially characterized  by the symmetries of the
tree approximation, i.e.~the ST identity (\ref{STG}) and
the Ward identities of rigid symmetry (\ref{wi}).
To be more specific we want to outline the one-loop construction
of the CS equation of the SM according to these
symmetries.

Applying the quantum action principle \cite{LOW71, LAM73}  one derives 
from (\ref{cstree}) that the dilatations
in 1-loop order are  broken by
\begin{equation}
\label{csbreak}
\biggl( m\partial _m -  \intd v \Bigl( {\delta \over \delta H } +
\zeta {\delta \over \delta \hat H } - \alpha_{inv}
 {\delta \over \delta \hat \varphi_o } 
\Bigr) \biggr)\Gamma =
\Delta_m + O(\hbar^ 2) \; ,
\end{equation}
where $\Delta_m$ is an integrated
field polynomial in quantum and external fields compatible with
 ultraviolet dimension 4 and infrared dimension 2,
neutral with respect to electric and ghost charge and CP-even.
According to the fact that the l.h.s is BRS-symmetric and symmetric
with respect to rigid symmetry (\ref{wsymsoft}), one gets
\begin{equation}
\label{breaksym}
s_{\Gamma_{cl}} \Delta_m = 0,  \qquad
{\cal W}_\a \Delta_m = 0 . 
\end{equation}
Thereby $s_{\Gamma_{cl}} $ is the linearized version of the ST operator
and acts on the quantum fields in the same way
as the classical BRS transformations.
We have therefore the task to find all independent field polynomials
satisfying the above constraints and to express them in form of
symmetric  differential operators. 

The 2- and 3-dimensional polynomials are already exhausted in the
l.h.s, if one takes higher-order corrections of the shift and the
parameter $\alpha _{inv} $ into account.
The 4-dimen\-sional polynomials are classified according to 
BRS variations, which are related to the anomalous dimensions,
and non-variations, which are related to the $\beta$-functions.
First we give a list of all symmetric field operators in the vector-ghost
sector:
\begin{eqnarray}
{\cal N}_V &=& \intd \Bigl( V_a {\delta \over \delta V_a} -
                        \rho_\alpha    {\delta \over \delta \rho_\alpha }
                       \nonumber \\
  & & +
                      \frac 1 \cwg (\sg c_Z + \cw c_A ) 
                       \bigl( \sw {\delta \over \delta c_Z   } 
                              +\cg{\delta \over \delta c_A   } \bigr)\Bigr)
\nonumber\\  
\hat {\cal N}_V & = & \intd \Bigl( (\sw Z + \cw A ) \bigl( \sw
                       {\delta \over \delta Z } + 
                       \cw {\delta \over \delta A    } \bigr) \nonumber \\
  & & +                      \frac 1 \cwg (\sg c_Z + \cw c_A )
                     \bigl( \sw {\delta \over \delta c_Z   }                 
                       +\cg{\delta \over \delta c_A   } \bigr)\Bigr)\nonumber
\\
\label{vec}
{\cal N}_B & = & \intd \Bigl( B_a {\delta \over \delta B_a} +
\bar c_a {\delta \over \delta \bar c_a}    
                           \Bigr) \\
\hat{\cal N}_B & = & \intd \Bigl(    (\sw B_Z + \cw B_ A ) \bigl( \sw
                       {\delta \over \delta B_Z } + 
                       \cw {\delta \over \delta B_A    } \bigr)\nonumber
  \\
  & & +                      \frac 1 \cwg (\sw \bar c_Z + \cg \bar 
                           c_A )
                     \bigl( \sg {\delta \over \delta \bar c_Z   }
                       +\cw{\delta \over \delta \bar c_A   } \bigr)\Bigr)
 \nonumber\\   
{\cal N}_c & = & \intd \Bigl(c _+ {\delta \over \delta c_+   }      +
                           c _- {\delta \over \delta c_-   }      
                          \nonumber\\
             & & +        \frac 1 \cwg (\cg c_Z - \sw c_A )
                     \bigl( \cw {\delta \over \delta c_Z   }                 
                       -\sg{\delta \over \delta c_A   } \bigr)\nonumber\\   
             & & -           \sigma _+ {\delta \over \delta \sigma_+   }      
                 -          \sigma _- {\delta \over \delta \sigma_-   }      
                    -  \sigma_3    {\delta \over \delta \sigma _3   } 
     \nonumber
\Bigr) .
\end{eqnarray}

Some remarks are in order concerning the special form of these
operators: According to rigid symmetry the counting operators of
the charged fields are related to the ones of the neutral sector
restricting the number of independent operators  to two for the vectors
and $B_a$~fields and one for the ghosts. Invariance under the ST identity
relates the abelian field differential operators of ghosts and
vectors. Furthermore it is seen that the 
abelian operator is not a BRS-variation and
is related to the $\beta$-functions by the local abelian
Ward identity (\ref{wardloc}), as it is usual in abelian gauge theories. 
The respective relation  is  given in (\ref{abrel}). 
 For completeness we have given
the field  operators for arbitrary ghost mass ratio
(\ref{gmr}); if one has introduced
normalization conditions which set the ghost angle equal to the 
weak mixing angle, $\theta_G$ can be immediately replaced by $\theta_W$
in the above expressions.

The symmetric field operators of  fermions can be split
into the ones of the left-handed and right-handed fields. Due
to the fact that we do not consider fermion mixing, we do not
have to consider mixed operators between different fermion families:
\begin{eqnarray}
\label{ferm}
{\cal N}^L_{F_{\delta,i}}& =& \intd \Bigl( \overline {F^L_{\delta,i}}
 {\delta \over \delta 
                         \overline {F^L_{\delta,i}}} 
                      -\overline {\Psi^R_{\delta,i}} {\delta \over \delta 
                         \overline {\Psi^R_{\delta,i}}} +
                     {\delta \over \delta    {F^L_{\delta,i}}}     {F^L_{
\delta,i}} 
-                     {\delta \over \delta    {\Psi^R_{\delta,i}}}  
{\Psi^R_{\delta,i}} 
\Bigr), \quad \delta= l,q    \\
{\cal N }^R_{f_i} & = & \intd \Bigl(  \overline {f^R_{i}} {\delta \over \delta 
                         \overline {f^R_{i}}} 
                      -\overline {\psi^L_{f,i}} {\delta \over \delta 
                         \overline {\psi^L_{f,i}}} +
                     {\delta \over \delta    {f^R_{i}}}     {f^R_{i}} 
-                     {\delta \over \delta    {\psi^L_{f,i}}}  {\psi^L_{f,i}} 
\Bigr),  \qquad f_i = e_i, d_i , u_i . \nonumber 
\end{eqnarray}

The  field operators of scalars comprise also the ones of the
external scalars. They are symmetric with respect to the rigid
operators, if one includes the shift of the Higgs field and the
external Higgs field:
\begin{eqnarray}
\label{scal}
{\cal N}_S +v \intd {\delta \over \delta H} & = &  \intd \Bigl( 
        ( \Phi  + {\mathrm v} ) ^\dagger
                        {\delta \over \delta \Phi^\dagger} + 
               {\delta \over \delta \Phi} ( \Phi + \mathrm v) 
            -  Y^\dagger {\delta \over \delta Y^\dagger}
                  - {\delta \over \delta Y} Y \Bigr) \\
{\cal N}_{\hat  S} +\zeta v \intd {\delta \over \delta \hat H} & = &
  \intd \Bigl( 
        ( \hat \Phi  + \zeta {\mathrm v} ) ^\dagger
                        {\delta \over \delta \hat \Phi^\dagger} + 
               {\delta \over \delta \hat \Phi} ( \hat \Phi + \zeta
               \mathrm v)  
    +  q^\dagger {\delta \over \delta q^\dagger}
                 + {\delta \over \delta q} q \Bigr) . \nonumber 
\end{eqnarray}
Among the symmetric insertions there is one which mixes the external
scalar with the quantum scalar. Defining the following
mixed operator, which is symmetric with respect to rigid transformations,
 \begin{eqnarray}
 {\tilde {\cal N}}_S +\zeta v \intd {\delta \over \delta H} & = &  \intd \Bigl( 
        ( \hat \Phi  + \zeta {\mathrm v} ) ^\dagger
                        {\delta \over \delta \Phi^\dagger} + 
               {\delta \over \delta \Phi} ( \hat \Phi + \zeta
                  \mathrm v)  \Bigr) ,
\end{eqnarray}
the corresponding BRS-symmetric  insertion is given by:
\begin{equation}
\label{extscal}
 \left( 
          {\tilde {\cal N}}_S +\zeta v \intd {\delta \over \delta H}\right)
              \Gamma_{cl} +
          \intd (  {\mathrm q}^\dagger Y - Y^\dagger {\mathrm q}) .
\end{equation}

The remaining symmetric insertions have to be generated by differentiating
the classical action with respect to the free parameters; these are
the  coupling $e$, which is the perturbative expansion parameter,
and furthermore the mass ratios, $\frac {M_W}{M_Z}$, for the weak
interactions, $\frac{m_H}{M_Z}$ for the scalar interaction, and
$\frac {m_{f_i}}{M_Z}$ for the Yukawa interactions. At this stage it
is unavoidable to treat $\theta _G$, i.e.~the ghost mass ratio
(\ref{gmr}), as
an independent parameter, because its differentiation corresponds
to an independent insertion in the gauge-fixing and ghost sector.
Similarly it turns out that also the differentiations with respect
to the two gauge parameters $\xi$ and $\hat \xi$ have to be included
(cf.~(\ref{fixext})). 

 Differentiations with respect to parameters which do not
appear in the ST identity and the rigid Ward operators 
of the tree approximation directly correspond to symmetric insertions:
\begin{equation}
\label{symcoup}
  {m_{H}} \partial _{ m_{H}} , \,
 {m_{f_i}} \partial _ {m_{f_i}} , \, \xi \partial_\xi,
\, \hat\xi \partial _{\hat \xi} \; .
\end{equation}
The differentiation with respect to the coupling $e$ is immediately 
symmetrized if one includes the shift; the operator
\begin{equation}
\label{coup}
e \partial_e - e \partial _e v \intd \Bigl( {\delta \over \delta H } +
\zeta {\delta \over \delta \hat H} \Bigr)  =
e \partial_e + \frac 2 e M_Z \sw \cw \intd \Bigl( {\delta \over \delta H } 
+ \zeta {\delta \over \delta \hat H} \Bigr) 
\end{equation}
is BRS- and rigid symmetric.
However, the differentiations with respect to the weak mixing angle  and to
the ghost angle
\begin{equation}
\partial_{\theta_W} = - M_Z \sw \partial_ {M_{W}}
 ,\quad
 \partial _
{\theta_G}
\end{equation}
have to be supplemented by field differentiations in order to be
BRS-symmetric:
\begin{eqnarray}
\label{thetaw}
\tilde \partial _{\theta_W} & \equiv&
\partial_{\theta_W} +  \intd\Bigl( A {\delta \over \delta Z} - 
                            Z{\delta \over \delta A}  
               +         B_ A {\delta \over \delta B_Z} - 
                          B_  Z{\delta \over \delta B_A} \Bigr) \nonumber \\   
               & &       + \frac 1 \cwg  \intd 
 c_ A \Bigl( {\delta \over \delta c_Z} + \swg 
                            {\delta \over \delta c_A} \Bigr)
\nonumber \\
 & &-  \frac 1 \cwg \intd \Bigl(  \bar
 c_ Z + \swg \bar c_A \Bigr) {\delta \over \delta\bar c_A } \\ 
\tilde \partial_{\theta_G} &\equiv &
\partial_{\theta_G}   -  \frac 1 \cwg  \intd 
 c_ Z \Bigl( {\delta \over \delta c_A} + \swg 
                            {\delta \over \delta c_Z}  \Bigr)
\nonumber \\
\label{thetag}
& & +  \frac 1 \cwg \intd \Bigl( \swg \bar
 c_ Z + \bar c_A \Bigr) {\delta \over \delta\bar c_Z} .
\end{eqnarray}
These two operators are immediately symmetric with respect to
the rigid transformations up to soft insertions corresponding
to the shift. The shift depends in the tree approximation on
the weak mixing angle and one has to enlarge $\tilde \partial _{\theta_W}$
by the differentiation with respect to the Higgs field and the
external Higgs field.
The operator
\begin{equation}
\label{weinsym}
\tilde \partial _{\theta_W} - \partial _{\theta_W} v \intd \Bigl(
{\delta \over \delta H} + \zeta {\delta \over \delta \hat H} \Bigr) = 
\tilde \partial _{\theta_W} - \frac 2e M_Z \cos 2\theta_W  \intd \Bigl(
{\delta \over \delta H} + \zeta {\delta \over \delta \hat H} \Bigr) 
\end{equation}
is then also rigid symmetric.

Acting with the symmetric operators 
(\ref{vec}), (\ref{ferm}), (\ref{scal}), (\ref{symcoup}), (\ref{coup}),
(\ref{weinsym}) and (\ref{thetag}) on the classical action one produces
together with the polynomial (\ref{extscal})
a complete basis for the breaking of the symmetric dilatational operator
(\ref{csbreak}) in 1-loop order. Therefore it is possible to give
the breaking in the form of a CS equation, i.e.~as a linear combination
of differential operators. Writing all the soft breakings produced 
by symmetrization with respect to the shift on the r.h.s we get
in 1-loop order the CS equation of the SM:
\begin{eqnarray}
\label{cs}
\! &\! \Bigl( \! & \! 
m\partial _m + \beta_e e\partial_e - \beta_{M_W} \tilde \partial _{\theta_W}
+ \beta_{m_H} m_H \partial _{m_H} + 
\sum_{i=1}^{N_F}\sum_f\beta _{m_{f_i}}  {m_{f_i}} \partial
_{m_{f_i}} - \beta_{\theta_G} \tilde \partial _{\theta_G}   \\
\! & \! -\! & \! \gamma_V {\cal N}_V  
-\gamma_B {\cal N}_B -\gamma_\xi  \partial_\xi 
- \gamma_c {\cal N}_c 
-\hat \gamma_V \hat {\cal N}_V
-\hat \gamma_B \hat {\cal N}_B - \gamma_{\hat \xi} \partial_{\hat \xi} 
- \gamma_S {\cal N}_S 
- \gamma_{\hat S} {\cal N}_{\hat S}
 - \tilde \gamma_S \tilde {\cal N}_S  \nonumber \\
\! & \!-\! & \! \sum_{i=1}^{N_F} \bigl( \gamma_{F_{l,i}}  {\cal N}^L_{F_{l,i}}
+\gamma_{F_{q,i}}  {\cal N}^L_{F_{q,i}}
+ \gamma_{e_{i}}  {\cal N}^R_{e_{i}}
+ \gamma_{u_{i}}  {\cal N}^R_{u_{i}}
+ \gamma_{d_{i}}  {\cal N}^R_{d_{i}} \bigr)  \,
\Bigr) \, \Gamma   \nonumber  \\
\! & \! = \! & \! \intd \Bigl( (1 + 
 \beta_e e\partial_e  - \beta_{M_W}\partial_{\theta_W}) v
\Bigl( {\delta \Gamma \over \delta H } +
\zeta {\delta \Gamma \over \delta \hat H } \Bigr)  
 +  v (\gamma_S +\tilde \gamma_S ){\delta \Gamma \over \delta H } +
 \zeta v \gamma_{\hat  S}{\delta \Gamma \over \delta \hat H } +
\alpha_{inv}
{\delta \Gamma \over \delta \hat \varphi_o }  \Bigr) \nonumber \\
\!& \! + \! & \! \intd \tilde \gamma_S \bigl(q^\dagger Y - Y^\dagger q \bigr) .
\nonumber 
\end{eqnarray}

Further information on the coefficient functions can be  achieved by
using the local Ward identity (\ref{wardloc}), which expresses
gauge invariance of the classical action under the abelian transformation
(\ref{GMN}).
Calculating the commutator of the CS operator and the local
Ward operator one gets:
\begin{equation}
\label{abrel}
\beta_e = \frac {\sw}{\cw} \beta_{M_W} + \gamma_V + \hat \gamma_V \; .
\end{equation}

Since  we have used a linear gauge fixing in the propagating fields
all the Green functions  which include  B-fields in the external
legs do not
get logarithmic higher-order corrections. The action of 
$m\partial_m $ on these Green functions  is therefore trivial  according
to their canonical dimensions.  Therefrom we get:
\begin{equation}
\label{garel}
\begin{array}{cclcccl}
\gamma_B & = & - \gamma_V &\quad&  \hat  \gamma_B & = & - \hat \gamma _V \\
\gamma_\xi & = & 2 \xi \gamma_V  
& \quad & \gamma_{\hat \xi} & = &
  2 (\gamma_V + \hat \gamma _V) \hat \xi + 2\xi \hat \gamma_V\\
\beta_{\theta_G} & = & \sg \cg \gamma_V &\quad &
\gamma_{\hat S} &= & \beta_e + \frac {\cw}{\sw} \beta_{M_W} + 
\gamma_V - \gamma_S  \; .
\end{array}
\end{equation}
From the explicit 1-loop expressions it is seen that the choice 
$\theta_G = \theta_W$ and $\hat \xi =0 $
(\ref{gmin})  is not stable under renormalization:  Since the
coefficients of the respective differential operators 
$\beta_{\theta_G} $ and $\gamma_{\hat \xi}$ are functions of 
 the anomalous dimensions of vectors and for this reason non-vanishing
(see (\ref{ga_V})),
 the  differentiation with respect to the independent
parameters $\hat \xi$
and $\theta _G$ has to be included in order to be able
to formulate the CS equation.

According to the derivation, the CS equation (\ref{cs}) and the
relations (\ref{abrel}) and (\ref{garel}) are
  valid in this form only in 1-loop
order. Proceeding to higher orders of perturbation theory 
goes along the same lines as in the 1-loop order, especially
the number of independent operators remains the same
to all orders of perturbation theory.
Their explicit form 
is modified order by order
in such a way that they become symmetric
with respect
to the ST identity and rigid Ward identities valid for higher-order
Green functions. Modifications essentially arise from establishing
the normalization conditions for separating massless/massive
fields at $p^2 = 0$ in addition to the usual on-shell conditions
for the masses.
It is interesting to note that the importance of these normalization
conditions for the off-shell existence of Green functions
can be already seen from the CS equation: Insertions with 
 infrared degree 2 as 
 $A^\mu A_\mu$
 on the r.h.s.\ are not forbidden from
infrared power counting, but those terms are mass terms for massless
particles and endanger infrared existence off-shell
and physical interpretation.
 In particular
 insertions of such field polynomials in higher-order
Green functions are non-integrable  and have to be proven to be absent.
(A similar analysis has been carried out in a simple non-gauge model
with spontaneously broken symmetry in \citere{CalSym}.)
The test with respect to the respective 2-point functions
at $p^2 = 0$ shows that these terms vanish if the mass matrix
of massless/massive fields is diagonal at $p^2 = 0$.

{\samepage
\newsection{The 1-loop coefficient functions}
Due to the spontaneous symmetry breaking mechanism the CS equation
has an unconventional form compared to the symmetric $SU(2) \times
U(1)$ gauge theory: 
}
It is an inhomogeneous equation with a
soft mass insertion on the r.h.s., because dilatations are broken by
the mass terms. The hard anomalies have to be absorbed in $\beta$-functions
with respect to mass differentiations and cannot be expanded in power
series of couplings according to the loop expansion. 
Moreover, since the $W$- and $Z$-bosons have different 
masses the anomalous dimensions are not purely leg counting operators
in the neutral sector, but include field operators, which mix the neutral
vectors and ghosts. Such field operators are not present in the
renormalization group equation of the symmetric theory. 
The perturbative 
expansion parameter is the electromagnetic coupling, which gives rise
to a  $\beta$-function $\beta_e$. 

We want to demonstrate how the $\beta$-functions and anomalous dimensions
are determined from the CS equation by testing with respect to 
appropriate vertices.  For this purpose we give
that part of the CS equation which is relevant for the test
with respect to vectors, quantum scalars and fermions
in its explicit form, setting
external fields and ghosts to zero: 
\begin{eqnarray}
\label{csexp}
&\hspace{-3mm} &
 \hspace{-3mm}\biggl\{
 m \partial _{ m} +\beta_e e\pa_e +
\beta _{m_H}m_H \pa _{m_H}  + \sum_{f_i}\beta _{m_{f_i}} 
m_{f_i}\pa _{m_{f_i}}   
- \beta_{ {M_W}}
\Bigl(   \pa_{\theta _W}
-   \intd\! \bigl(
Z^\mu {\delta \over \delta A^\mu } 
 - A^\mu {\delta \over \delta Z^\mu } \bigr) 
\Bigr)  
\nonumber \\
&\hspace{-3mm} &\hspace{-3mm} - \gamma_V \Bigl( \intd \!
V^\mu_a {\delta \over \delta V_a^\mu } 
 + 2 \xi \partial _\xi + 2 \xi \partial_{\hat \xi} + \sg \cg \pa_{\theta_G}
\Bigr) 
- \gamma _S\intd\! 
 \bigl( \Phi{\delta \over \delta \Phi } + {\delta \over \delta \Phi^\dagger
 } \Phi^\dagger \bigr) 
 \\
&\hspace{-3mm}&\hspace{-3mm} - \hat\gamma^{V} \Bigl( \intd \!(
 {\sin \theta_W}
 Z^\mu  + \cw A^\mu   )
 \bigl( \sw {\delta \over \delta Z^\mu } + \cw {\delta \over \delta A^\mu }  
\bigr)
+ 2 (\hat \xi + \xi )\pa _{\hat \xi} \Bigr)\nonumber \\
&\hspace{-3mm}&\hspace{-3mm} - \sum_{F_{\delta,i}}\gamma _{F_{\delta,i}}
\intd\!  \Bigl( \overline {F^L_{\delta,i}}
 {\delta \over \delta   \overline {F^L_{\delta,i}}} 
          +  {\delta \over \delta    {F^L_{\delta,i}}}     {F^L_{
\delta,i}} \Bigr)
- \sum_{f_i} 
\gamma_{f_i} \intd \! \Bigl(  \overline {f^R_{i}} {\delta \over \delta 
                         \overline {f^R_{i}}}  +
{\delta \over \delta    {f^R_{i}}}     {f^R_{i}}  \Bigr) 
  \biggr\} \Gamma \bigg|_{ext.f.= 0 \atop c_a,\bar c_a =0} \nonumber
\\
&\hspace{-3mm}&\hspace{3mm} = [\Delta_s]_3 ^3 \cdot \Gamma . \nonumber
\end{eqnarray}
One is able to determine the $\beta$-functions from vertex functions of
UV-dimension 4, using thereby that the soft insertion will
vanish for asymptotic Euclidean momenta much larger than the mass
of the heaviest particle involved. Similarly the anomalous dimensions
are determined from the residua at asymptotic momentum.
We evaluate the coefficient functions in the most general linear 
gauge fixing invariant with respect to rigid symmetry transformations
as given in \refeq{fixext}. For completeness, in the appendix
we list the
propagators of vector and scalar fields in the general linear
$R_\xi$ gauge (\ref{eq:gf}). From there the propagators compatible with rigid
symmetry are obtained by assigning to the gauge-fixing parameters the values
(\ref{xidef}) -- (\ref{zetadef3}). 
In the computation of the 
CS coefficient functions only those parts of propagators 
of vector and scalar fields contribute
which behave like $1 \over p^2$ in the expansion for
asymptotically large $p^2$.

First we  determine the 
$\beta$-functions of the electromagnetic coupling, $\beta_e$,
and of the $W$-boson mass, $\beta_{M_W}$,
and the anomalous dimensions of the vectors $\gamma_V$
and $\hat \gamma_V$. The anomalous dimensions of the
vectors are calculated from the transverse parts of the 2-point functions:
\begin{equation}
\Gamma^{\mu\nu}_{V_a V_b} (p) \equiv - \left( \eta ^{\mu \nu} - 
{p^\mu p^\mu \over p^2 } \right) \Gamma^T_{ab}(p^2)
- {p^\mu p^\mu \over p^2 }  \Gamma^L_{ab}(p^2) .
\end{equation}
In the tree approximation we have 
\begin{equation} 
\label{2tree}
\partial _{p^2}\Gamma^{T(0)}_{ab} = \tilde I_{ab} .
\end{equation}
We find therefore in the asymptotic region
(all functions involved are purely of 1-loop order)
\begin{eqnarray}
\label{eq:gammas}
m\partial_m \partial_{p^2}\Gamma^{T(1)}_{+-} &
{\buildrel p^2\to -\infty \over 
=}&  2\gamma_V^{(1)} \\
m\partial_m \partial_{p^2}\Gamma^{T(1)}
_{ZZ} &{\buildrel p^2\to -\infty \over =}&
  2 (\gamma_V^{(1)} + \sws \hat \gamma_V^{(1)})  \nonumber \\
m\partial_m \partial_{p^2}\Gamma^{T(1)}
_{AA} &{\buildrel p^2\to -\infty \over =}&
  2 (\gamma_V^{(1)}  + \cws \hat\gamma_V^{(1)}) \nonumber \\
m\partial_m \partial_{p^2}\Gamma^{T(1)}
_{ZA} &{\buildrel p^2\to -\infty \over =}&
  2 \sw \cw \hat\gamma_V^{(1)} , \nonumber 
\end{eqnarray}
which determines the high-energy logarithms of the one-loop
self-energies,
e.g.\ 
\begin{equation}
\left(\partial_{p^2}\Gamma^{T(1)}_{+-} \right)_{\mathrm lead.\ log} =
- \gamma^{(1)}_V \ln \frac{|p^2|}{m^2}.
\end{equation}
Accordingly, 
the anomalous dimensions $\gamma_V$ and $\hat \gamma_V$ are obtained by
calculating the high-energy logarithms of two of the self-energies
appearing in \refeq{eq:gammas}, while the leading logarithms of the
other two self-energies are then already fixed.
In the general linear gauge specified in \refeq{fixext}
we get the following result:
\begin{eqnarray}
\label{ga_V}
\gamma_V^{(1)} &=& {e^2 \over 4 \pi^2 \sin^2 \theta_W}
\Bigl( {6 \xi -25 \over 24 } 
                         + {1 \over 3 } N_F \Bigr) \\
\hat \gamma_V ^{(1)}&=&  {e^2 \over 4 \pi^2 }
 \Bigl( - {6 \xi -25 \over 24  \sin^2 \theta_W} 
                             + {1 \over 24  \cos^2  \theta_W} 
                             +{-3 + 8 \sin^2 \theta_W \over 9  
                                \sin^2 \theta_W \cos^2 \theta_W}
 N_F \Bigr) .   
\end{eqnarray}
The fermion contributions to the coefficients of the high-energy
logarithms are the same as those given in \citere{bfmlong} in the 
framework of the background-field method, while the contributions
of the vector bosons are gauge-parameter-dependent.
The gauge-parameter dependence of the anomalous dimensions is the same
as in usual $R_{\xi}$~gauges, i.e.\ they are
independent of the abelian gauge parameter
$\hat \xi$, the ghost mass parameter $\zeta$ and  of
the ghost angle (\ref{gmr}). This is seen most easily by noting that
the diagrams which contribute to the photon self-energy and  to
the Z-photon self-energy have only charged fields in internal lines and
are not affected by non-diagonal propagators in the neutral sector nor
by transformations of neutral ghosts (\ref{gred}).
 As before $N_F$ denotes the number of fermion generations.

The $\beta$-function $\beta_{M_W}$ can be determined from the
neutrino--neutrino--photon vertex at high energies.
Testing the CS equation (\ref{csexp}) with respect to this
vertex we get the following result:
\begin{equation}
m \partial _m 
\Gamma_{\bar \nu \nu A_\mu} ^{(1)}
{\buildrel p^2\to -\infty \over =} (\beta_{M_W}^{(1)}
 + \sw \cw \hat \gamma_V ^{(1)})
\Gamma^{(0)}_{\bar \nu \nu Z_\mu}\, . 
\end{equation}
Since the photon does not couple to neutrinos in the tree approximation
the contributions on the r.h.s.\ completely
  arise from the mixed field operators
$A {\delta \over \delta Z}$ present in the symmetric operators of
$\beta^{(1)}_{\theta_W}$ and $\hat \gamma_V^{(1)}$. 
In the Feynman gauge ($\xi = 1 , \hat \xi = 0$) one has
\begin{equation}
\beta^{(1)} _{M_W}   + \sw \cw \hat \gamma_V^{(1)}  = - {e^2\over 4 \pi^2}
 {\cw\over \sw} ,
\end{equation}
which gives the result
\begin{equation}
\label{betaMW}
\beta_{M _W}^{(1)} = - {e^2 \over 4\cdot 24 \pi^2\sw \cw} 
                     \Bigl( (43 - 8 N_F) - (42 - 
                     \hbox{$\frac {64}3$} N_F ) \sin^2 \theta_W \Bigr) .
\end{equation}
Applying the algebraic control of gauge-parameter dependence~\cite{PiSi85} to
spontaneously broken theories~\cite{HaKr96} 
it can be derived that $\beta_{M_W}$ is gauge-parameter-independent.

The abelian relation (\ref{abrel}) allows to determine $\beta^{(1)}_e$
without calculating further diagrams 
from the results obtained for $\gamma^{(1)}_V$, $\hat \gamma^{(1)}_V$ and
$\beta^{(1)}_{M_W}$:
\begin{equation}
\beta^{(1)}_e = -{e^2 \over 24\cdot4 \pi^2} 
\Bigl( 42 - \hbox{$\frac{64}3$} N_F \Bigr) .
\label{eq:betae}
\end{equation}
Alternatively $\beta^{(1)}_e$ can of course also directly be obtained from the
$W^+W^-A$~vertex at asymptotic momenta
\begin{equation}
\label{wwa}
 m\partial_m \Ga_{W_+^\mu W_-^\nu  A^\rho}^{(1)} \longrightarrow 
(3 \ga_V^{(1)} - \beta_e^{(1)} - {\cw \over \sw} \beta_{M_W}^{(1)})
\Ga^{(0)}_{W_+^\mu W_-^\nu  A^\rho} 
\end{equation}
 or from the $\bar e e A$~vertex (cf.~(\ref{eq:eegavert})).
 We have explicitly
checked that this indeed results in $\beta_e^{(1)}$ as given in
\refeq{eq:betae}. As can be seen in \refeq{eq:betae}, the
$\beta$-function of the electromagnetic coupling is QED-like in the
sense that it only depends on the electromagnetic coupling $e^2$ but not
on $\sin^2 \theta_W$. However, due to non-abelian interactions 
of the photon with the $W$-bosons it receives contributions with
negative sign. This leads to the fact that
$\beta_e^{(1)}$ in the SM has a negative sign if one includes only one 
fermion family. 
In section 6.2 we compare 
it to the QED $\beta$-function.

We now turn to the anomalous
dimensions of the fermions, 
which are needed e.g.\ for the independent determination of 
$\beta^{(1)}_e$ from the electron--electron--photon vertex.
Splitting the fermion self-energy into left- and right-handed parts
and into the scalar mass contribution,
\begin{equation}
\Gamma_{\bar f f} = \ps - m _f + \ps \frac 12 (1- \gamma^5) \Sigma^L_f
 + \ps \frac 12 (1+ \gamma^5) \Sigma^R_f + m \Sigma^m _f \; ,
\end{equation}
one is able to calculate $\gamma_{F_{\delta,i}}$ from left-handed
and $\gamma_{f_i}$ from right-handed contributions:
\begin{eqnarray}
m\partial_m \Sigma^{L(1)} _{e_i} &{\buildrel p^2\to -\infty \over =}&
 2 \gamma^{(1)}_{F_{l,i}} \; , \quad
m\partial_m \Sigma^{L(1)} _{d_i} \;\; {\buildrel p^2\to -\infty \over =} \;\;
 2 \gamma^{(1)}_{F_{q,i}}  \; ,\\
m\partial_m \Sigma^{R(1)} _{f_i} & {\buildrel p^2\to -\infty \over =} & 
2 \gamma^{(1)}_{f_i} \; .\nonumber
\end{eqnarray}
Calculating the high-energy logarithms of the fermion self-energy
contributions 
one gets the following result:
\begin{eqnarray}
\label{eq:gammaferm}
\gamma^{(1)} _{F_{l,i}} & = & \fsc 
           {1\over \sin^2 2\theta _W }
        \left[ \left(3 - 2 \sws\right) \xi + \sws \hat\xi +
                       \frac{m_{e_i}^2 }{M_Z^2} \right] \\
\gamma^{(1)} _{F_{q,i}} & = & \fsc 
                   {1\over \sin^2 2 \theta_W} 
        \left[\left(3 - \frac{26}{9} \sws \right) \xi + 
              \frac{1}{9} \sws \hat\xi+
              \frac{m_{u_i}^2 + m_{d_i}^2}{M_Z^2} \right] \\
 \gamma^{(1)} _{f_{i}} & = & \fsc 
                       {2 \over \sin^2 2 \theta_W} 
        \left[2 Q_f^2\sws \left(\xi + \hat\xi\right) +
                       \frac{m_{f_i}^2 }{M_Z^2} \right] . 
\end{eqnarray}
For the asymptotic behavior of the electron--electron--photon vertex we
obtain: 
\begin{eqnarray}
\label{eq:eegavert}
m \partial _m 
\Gamma^{(1)}_{\bar e e A_\mu} 
&{\buildrel p^2\to -\infty \over =} & \phantom{+} e
\Bigl(\gamma^{(1)}_V + \hat \gamma^{(1)}_V -\beta^{(1)}_e + 2 \gamma
^{(1)}_ e
 + \beta^{(1)}_{M_W} \frac 
{\sw}{\cw}  \Bigr)  \gamma^\mu\\
 &  & + e \Bigl(2 \gamma^{(1)}_{F_l} - 2 \gamma^{(1)}_e  - 
\frac 1 {\sin 2\theta_W}
(\beta^{(1)}_{M_W} + \sw \cw \hat \gamma^{(1)}_V )
 \Bigr) \gamma^\mu  \frac{1}{2}
(1-\gamma^5) .
\nonumber
\end{eqnarray}
In this formula we have already inserted the tree vertices 
\begin{eqnarray}
\Gamma^{(0)}_{\bar e e A_\mu}  & = & e\gamma^\mu   \\
\Gamma^{(0)}_{\bar e e Z_\mu} & = & - e \frac 1{\sin 2 \theta_W}
 \gamma ^\mu \frac{1}{2}
(1 - \gamma^5) + e\frac \sw \cw \gamma ^\mu . \nonumber
\end{eqnarray}  
As mentioned above, using \refeq{eq:eegavert} and 
the results of \refeq{eq:gammaferm} one can
check the abelian relation we have used to determine $\beta_e$.
It is seen in \refeq{eq:eegavert} 
that the parity non-violating contribution satisfies
an analogous relation as in QED: The high-energy logarithms of
the electron--electron--photon vertex are completely related to
the anomalous dimensions of (right-handed) electrons.
Due to the non-abelian
contributions there are however parity-violating high-energy logarithms
for the off-shell Green functions.

For calculating the remaining $\beta$-functions of fermion masses
and the Higgs mass one first has to determine the 
anomalous dimensions of the scalars.
We obtain
\begin{equation}
\gamma^{(1)}_S = \frac{e^2}{8 \pi^2} \frac{1}{\sin^2 2 \theta_W}
\left[\sum _i {m_{e_i}^2 +3 m_{d_i}^2 + 3 m_{u_i}^2 \over M_Z^2 } 
+ \frac{1}{2} \left(3 - 2 \sws\right) (\xi - 3) + \frac{1}{2} \sws \hat\xi
\right] .
\end{equation}
The $\beta$ functions $\beta^{(1)}_{m_{f_i}} $ and $ \beta^{(1)}_{m_H} $
are determined from the high-energy logarithms according to
the following formulas:
\begin{eqnarray}
m \partial _m 
\Gamma^{(1)}_{\bar f_i f_i H } 
&{\buildrel p^2\to -\infty \over =} & \Bigl( - \beta^{(1)}_e - 2
{\cos 2 \theta_W  \over \sin 2 \theta_W } \beta^{(1)}_{M_W} -
\beta^{(1)}_{m_{f_i}} + \gamma^{(1)}_S +  \gamma^{(1)}
_{F_{\delta,i}} +  \gamma^{(1)}_{f_i} \Bigr)
\Gamma^{(0)}_{\bar f_i f_i H}  \nonumber\\
m \partial _m 
\Gamma^{(1)}_{HHHH } &
{\buildrel p^2\to -\infty \over =}& 2 \Bigl( - \beta^{(1)}_e - 2
{\cos 2 \theta_W  \over \sin 2 \theta_W } \beta^{(1)}_{M_W} -
\beta^{(1)}_{m_{H}} + 2 \gamma^{(1)}_S  \Bigr)
\Gamma^{(0)}_{HHHH} \; .
\end{eqnarray}
Therefrom we derive the result:
\begin{eqnarray}
  \beta^{(1)}_{m_{e_i}} \is {e^2 \over 24 \pi^2 \sin^2 2 \theta_W}  
 \Bigl\{ \frac 9 2 \frac {m_{e_i}^2}{M_Z^2} + 3 
         \sum_{f_j}  \frac {m_{e_j}^2 + 3 m_{u_j}^2
+ 3 m_{d_j}^2}{M_Z^2} 
  \\
& &\qquad
 + (\frac{59}2 - 8 N_F) - (95 - 16 N_F) \sws  + (42 - 64 \frac {N_F} 3)
\sin ^4\theta _W
\Bigr\} \nonumber\\
  \beta^{(1)}_{m_{u_i}} \is {e^2 \over 24 \pi^2 \sin^2 2 \theta_W}  
 \Bigl\{ \frac 92  \frac {m_{u_i}^2 - {m_{d_i}^2 }}{M_Z^2} + 3 
         \sum_{f_j} \frac {m_{e_j}^2 + 3 m_{u_j}^2
+ 3 m_{d_j}^2}{M_Z^2} 
 \\
& & \qquad
+ (\frac{59}2 - 8 N_F) - (81 - 16 N_F) \sws  + (42 - 64 \frac {N_F} 3)
\sin ^4\theta _W
\Bigr\} \nonumber\\
  \beta^{(1)}_{m_{d_i}} \is {e^2 \over 24 \pi^2 \sin^2 2 \theta_W}  
 \Bigl\{ \frac 92 \frac {m_{d_i}^2 - {m_{u_i}^2 }}{M_Z^2} + 3 
         \sum_{f_j}\frac {m_{e_j}^2 + 3 m_{u_j}^2
+ 3 m_{d_j}^2}{M_Z^2} 
\\
& & \qquad
+ (\frac{59}2 - 8 N_F) - (75 - 16 N_F) \sws  + (42 - 64 \frac {N_F} 3)
\sin ^4\theta _W
\Bigr\} \nonumber \\
  \beta^{(1)}_{m_H} \is {e^2 \over 24 \pi^2 \sin^2 2 \theta_W}  
 \Bigl\{  9 \frac {m_H^2 }{M_Z^2} + 6 
         \sum_{f_j} \frac {m_{e_j}^2 + 3 m_{u_j}^2
+ 3 m_{d_j}^2}{M_Z^2}  \\
& & \qquad
  - 12         \sum_{f_j}  \frac {m_{e_j}^4 + 3m_{d_j}^4 + 3
m_{u_j}^4}{M_Z^2 m_H^2}   \nonumber \\
& & \qquad 
+ (16 - 8 N_F) - (68 - 16 N_F) \sws  + (42 - 64 \frac {N_F} 3)
\sin ^4\theta _W
 \nonumber\\
& & \qquad+{ M_Z^2 \over m_H^2} 
\Bigl( 27 -  36 \sws + 18 \sin^4 \theta_W \Bigr) \Bigr\} . \nonumber
\end{eqnarray}
{}From considerations of gauge-parameter dependence it is seen that
these $\beta$-functions are 
gauge-parameter-independent~\cite{PiSi85,HaKr96}. As mentioned above,
the same holds for the $\beta$-functions of the electromagnetic coupling 
and the vector-boson mass ratio.

Finally, by testing with respect to
the ghost self-energy one finds the following
result for the anomalous dimension of the ghosts, $\ga _c$:
\begin{equation}
\label{ga_c}
\ga^{(1)}_c =   \frac {e ^2} {4 \pi^ 2 \sin^2\theta_W } \left( 
\frac \xi 2 - \frac {43} { 24} + \frac 13 N_F \right) .
\end{equation}
In the Landau gauge ($\xi = 0$)
the anomalous dimension of the Faddeev-Popov ghosts
is equal to the $\beta$-function of the non-abelian gauge coupling 
$g_2$ in (\ref{eq:betrel4}):
\begin{equation}
\ga^{(1)}_c\Bigr|_{\xi = 0} = \beta_e ^{(1)} + 
{\cos \theta_W \over \sin \theta_W} 
\beta_{M_W}^{(1)} .
\end{equation}
 This
coincidence is not accidental, but is derived from  the existence of an
integrated
antighost equation in the Landau gauge. 

  

Having determined the 1-loop $\beta$-functions and anomalous
dimensions, it is possible to determine the high-energy logarithms
of any 1-loop vertex function of the Standard Model
in an analogous way as shown, for instance, in \refeq{wwa} for the
W-boson--photon vertex
and in (\ref{eq:eegavert}) for the electron--electron--photon  vertex. 
Since we have also calculated  the anomalous dimensions of ghosts
(\ref{ga_c}),
this is also possible for the external field vertices appearing in
(\ref{gaext}), which determine the
higher-order corrections to the BRS transformations.

In their applications the importance of the CS and RG equation is founded 
in the fact that from the knowledge of the equations at 1-loop order
one can draw conclusions for the asymptotic behavior of the
vertex functions in higher orders.
In particular,
if the 1-loop coefficient functions of the CS equation are given in
a general gauge as has been worked out above, one is able to determine 
   the quadratic (leading) logarithms of 
2-loop  order for any vertex function of the Standard Model.
 For illustration we  evaluate the CS equation for the photon
self-energy in 2-loop order at an asymptotically large momentum:
\begin{eqnarray}
\label{quadlogs}
m\partial_m \partial_{p^2} \Ga_{AA}^{T(2)}  \!
& \! \stackrel{p^2 \to - \infty} = \! & \!\Bigl( 2 \ga_V ^{(1)} + 
2 \hat \ga_V^{(1)} \cos^2 \theta _W 
- \beta_e ^{(1)} e \partial_e  + \beta_{M_W}^{(1)} \partial _{\theta_W} +
2 \ga_V^{(1)}\xi \partial_\xi \Bigr) 
\partial_{p^2} \Ga_{AA}^{T(1)} \nonumber \\  
\!&\! \! & \! + ( 2 \hat \ga_V^{(1)} \cos \theta _W \sin \theta_W  +
2 \beta_{M_W}^{(1)} )
\partial_{p^2} \Ga_{ZA}^{T(1)}  +  {\mathrm Const.}^{(2)} .
\end{eqnarray}
In ${\mathrm Const.}^{(2)}$
  all terms are included which approach constants
if we take the limit of asymptotically large Euclidean momentum $p^2$. 
They give rise to linear logarithmic contributions.
Contributions to these terms arise from three different sources:
\begin{enumerate}
\item Applying the CS equation (\ref{cs})
       with 2-loop coefficient functions to
      the tree vertices
  in analogy to the 1-loop case (\ref{eq:gammas}), one gets constant 
  contributions with 2-loop coefficient functions which have to be
  determined by testing with respect to appropriate vertex functions. 
  In the example above the 2-loop coefficient functions 
  read  $ 2 (\ga^{(2)} + \hat \ga ^{(2)} \cos^2 \theta _W)  $.
\item Since the  ST operator and the
Ward operators
of rigid symmetry 
 are renormalized   in the on-shell schemes, the symmetric
operators which build up the CS equation  (see section 4)
 get higher-order corrections. In 2-loop order these corrections
depend on the 1-loop corrections to the ST and Ward operators 
and on the CS coefficient functions of 1-loop order. 
 These 
contributions depend strongly on the normalization conditions, but can
be
determined by a 1-loop calculation.
\item In 2-loop order for asymptotic momenta not only logarithms arise
 from the 1-loop vertex functions, but also constants, i.e.\ in the example 
 above
\begin{equation}
\label{c_ZA}
\Ga_{ZA}^{T(1)} \stackrel{p^2 \to - \infty} =
 -  \sin \theta_W \cos \theta_W
\hat \ga^{(1)}_V  \ln \frac{|p^2|}{ m^2}  + C_{ZA}^{(1)} \,. 
\end{equation}
 The  finite constant $C_{ZA}^{(1)} $ 
is determined
 from the normalization conditions for diagonalizing the mass matrix of
photon and $Z$-boson on-shell and depends on the mass parameters of the
Standard Model.  In general  such finite constants as shown in the above
example 
are specific for on-shell normalization conditions  of spontaneously
broken
theories. 
\end{enumerate}
All these constant  terms contribute to the single logarithms of the photon
self-energy in 2-loop order and, of course,
the computation of
all single logarithms of  2-loop order demands a 
2-loop calculation. However, in the list above
 we have  given also such constant contributions
which can be determined from a  1-loop calculation. They all depend
strongly
on the normalization conditions. In \citere{InvCh} such
1-loop induced logarithmic contributions with large mass-dependent
logarithmic coefficients have been found
 in the  spontaneously broken Yukawa-Higgs model.
They arise from normalization-dependent 1-loop contributions as
discussed above and have to be separated from the mass-parameter
independent logarithmic contributions to the 2-loop order.
In \citere{InvCh} this has been achieved
 by using the consistency equation between  the CS equation and the 
RG equation. For further applications it is certainly very interesting 
 to completely single
out these large 1-loop induced contributions  
 by a self-consistent construction of higher-order solutions to the
CS equation  or by using the RG equation in a similar way as in
\citere{InvCh}.

Focusing now on the quadratic logarithms of 2-loop order,
we are able to evaluate (\ref{quadlogs}) by inserting the asymptotic
1-loop results (\ref{eq:gammas})
\begin{eqnarray}
\left(\partial_{p^2} \Ga_{AA}^{T(1)} \right) _{\mathrm lead.\ log}
& = & - (\ga_V^{(1)} + \cos^2 \theta_W
\hat \ga_V^{(1)} ) \ln \frac{|p^2|}{ m^2} \\
\left(\partial_{p^2} \Ga_{ZA}^{T(1)} \right) _{\mathrm lead.\ log} 
& = & -  \sin \theta_W \cos \theta_W
\hat \ga^{(1)}_V  \ln \frac{|p^2|}{ m^2} . \nonumber
\end{eqnarray}
  Further simplification
can be achieved by eliminating $\beta^{(1)}_e$ using the abelian relation
(\ref{abrel}) and by noting that the coefficient of the
leading logarithm of the
 photon self-energy depends only on the electromagnetic coupling in
1-loop order. 
Then we end up with
\begin{eqnarray}
\partial_{p^2} \Ga^{T (2)}_{AA} &\stackrel {p^2 \to -\infty} = & 
- \frac 12 (\hat \ga_V^{(1)} \sin \theta_W  \cos \theta_W + \beta^{(1)}_{M_W} )
\frac {\sin \theta_W}{\cos\theta_W} \ga^{(1)} _V \ln ^2  \frac{|p^2|}{ m^2} \\
&
& + \frac 12 \ga^{(1)}_V 
\xi \partial _\xi (\ga^{(1)}_V + \cos^2 \theta_W \hat 
\ga^{(1)}_V )
 \ln ^2  \frac{|p^2|}{ m^2} 
 + {\cal O} (\ln \frac{|p^2|}{ m^2} ) . \nonumber
\end{eqnarray}
Since the anomalous dimensions of  vectors are not
gauge-parameter-independent,
the derivative with respect to the gauge parameter contributes to
the 2-loop order leading logarithms and underlines the significance of
having calculated 1-loop coefficient functions in a general gauge.
Inserting the explicit expressions of 1-loop order, (\ref{ga_V})  
and (\ref{betaMW}), we finally get
\begin{eqnarray}
\label{eq:twoloopex}
\partial_{p^2} \Ga^{T (2)}_{AA} &\stackrel {p^2 \to -\infty} = &  
 \frac {e^2} {4 \pi ^2 } \ga_V^{(1)}  \ln ^2  \frac{|p^2|}{ m^2} \,
 \frac{1}{2} \left(\frac{\xi}{2} + \frac{3}{4} \right)
 + {\cal O} (\ln \frac{|p^2|}{ m^2} )  \\
&= & \frac {e^4} {16 \pi ^4 } \ln ^2  \frac{|p^2|}{ m^2} \,
 \frac{1}{2 \sws} \left(\frac{\xi}{2} + \frac{3}{4} \right)
 \left({6 \xi -25 \over 24 } + {1 \over 3 } N_F \right) 
+ {\cal O} (\ln \frac{|p^2|}{ m^2} ) . \nonumber
\end{eqnarray}
Here $m$ denotes the largest mass parameter of the Standard Model.
As a result, for asymptotic momenta much larger
than all masses of the theory the photon self-energy
 includes quadratic logarithms in 2-loop order.
This is in contrast to pure QED and
 is  caused by the non-abelian interaction of the photon with the
W-bosons and the Faddeev-Popov ghosts.

Further applications as well as a detailed consideration 
of the above-mentioned 1-loop induced large
logarithms in 2-loop order will be given elsewhere.

\newsection{Comparison with the massless symmetric theory and 
QED}
\newsubsection{Symmetric theory}

{}From pure power-counting arguments it has been reasoned that the
divergence structure of the symmetric theory is related to the
one of the corresponding spontaneously broken theory~\cite{SY72}.
 The divergence
structure corresponds  to the appearance of high-energy logarithms
order by order in perturbation theory. In 1-loop order the 
high-energy logarithms of the spontaneously broken theory are the same
as the ones of the symmetric theory,
since they arise  from diagrams with only
 4-dimensional vertices which are not affected
 by spontaneous breaking of the theory. The low-energy
structure of the spontaneously broken theory
is summarized in the r.h.s.~of the CS equation, which vanishes
if one goes to non-exceptional momenta much larger than the masses of 
the theory. As we have shown in the previous section, the
$\beta$-functions of the CS equation are related to the asymptotic
logarithms
of the model.
For this reason there exist in 1-loop order simple
relations between the $\beta$-functions of the electroweak Standard Model and
the corresponding $SU(2) \times U(1)$ massless symmetric theory.
These relations reflect the tree relations between the parameters
of the symmetric and the spontaneously broken theory. Since these
relations get higher-order corrections it is expected that the
higher-order $\beta$-functions get different higher-order contributions
in the spontaneously broken theory than in the symmetric one.
Indeed the constants which appear 
as a consequence of on-shell conditions in the asymptotic 
limit (see \refeq{c_ZA}) enter these higher-order corrections and
affect -- as already pointed out above -- the coefficients of 2-loop single
logarithms. A detailed computation of mass effects due to spontaneous
breaking of the theory and due to on-shell conditions has been carried out
in \citere{InvCh}.  In order to specify these
contributions one has to find  self-consistent solutions of the
CS equation in higher orders in the on-shell schemes.
In the present context we restrict ourselves to the 1-loop relations
between the $\beta$-functions of the symmetric and spontaneously broken theory.


As usually the independent parameters of the
massless symmetric theory are parameterized with the
$U(1)$-coupling $g_1 $, the $SU(2)$-coupling $g_2$, the Yukawa
coupling $G_{f_i}$ and the Higgs self-coupling $\lambda$.
With the conventions of \citere{Ara} the parameters
of the spontaneously broken theory are related to the couplings
of the symmetric theory in the tree approximation
 by
\begin{eqnarray}
\label{treerel}
g_1 & = & {e\over \cw} + {\cal O}(\hbar)\nonumber \\
g_2 & = & {e\over \sw} + {\cal O}(\hbar)\nonumber \\
{G_{f_i}} &= & {\sqrt 2  m_{f_i} \over M_Z \sin 2 \theta _W }+ {\cal O}(\hbar)\nonumber\\
\lambda &= & e ^2 { 4 m^2_H \over M^2_Z \sin^2 (2 \theta_W) } + {\cal
O}(\hbar).
\end{eqnarray}
 In the massless theory one has to introduce a scale parameter,
the normalization point $\kappa $, for fixing the coupling constants.
Usually there is introduced only one normalization point $\kappa$, its
variation $\kappa \partial _\kappa$ expresses at the same time
renormalization group invariance and breaking of dilatations.
The corresponding partial differential equation is then
valid to all orders of perturbation theory, 
\begin{eqnarray}
\label{cssym}
& &\biggl\{ \kappa \partial _\kappa + \beta_{g_1} \partial _{g_1}
+ \beta_{g_2} \partial _{g_2} 
+ \beta _\lambda \partial _\lambda  + 
\sum_{f_i}\beta_{G_{f_i}} \partial _{G_{f_i}} \\
& & -  \ga^V_1 \Bigl(\intd \! B^\mu {\delta \over \delta B^\mu}  -  \xi_1 
\partial _{ \xi_1} ) 
 - \ga ^V_2  \biggl(
\intd\! \Bigl( W^\mu_\a {\delta \over \delta W^\mu_\a} -
\bar c_\a {\delta \over \delta \bar c_\a } \Bigr) - \xi_2 \partial _{\xi_2} 
\biggr) \nonumber \\
& &  
- \sum_{i=1}^{N_F} \Bigl[ \sum_\delta 
\ga^L_{F _{\delta,i}} \intd\Bigl(\overline {F^ L_{\delta,i}} 
{\delta \over \delta \overline {F^ L_{\delta,i}} } +
{\delta \over \delta  {F^ L_{\delta,i}} } F^ L_{\delta,i} \Bigr)
-  \sum_f
\ga^R_{f_i} \intd\Bigl(\overline {f^ R_{i}} 
{\delta \over \delta \overline {f^ R_{i}} } +
{\delta \over \delta  {f^ R_{i}} } f^ R_{i} \Bigr) \Bigr] 
 \nonumber \\
& & 
- \ga ^S \intd\Bigl(\Phi {\delta \over \delta \Phi} + {\delta \over \delta 
\Phi^\dagger} \Phi^\dagger \Bigr) - \ga^g \intd c_\a 
{\delta \over \delta c_\a }  
\biggr\}\,   
\Ga \bigg|_{ext.f.= 0 }
= 0 .
\end{eqnarray}
Here  $W^\mu_\a ,\a = 1,2,3 $, are the $SU(2)$-gauge fields and
$B_\mu$ is the abelian gauge field. The abelian relation in the
symmetric theory reads:
\begin{equation}
\beta_{g_1} = \ga^ V_1 .
\end{equation}
Comparing (\ref{cssym}) with the CS equation of the Standard Model
 it is seen that
we did not have to introduce the external scalar doublet.
In addition we have left out the abelian ghosts,
 since they are free fields in the symmetric theory. 
Comparing the CS equation of the spontaneously broken theory
to the one of the symmetric theory for different high-energy
vertex functions one gets the following 1-loop relations by inserting
the tree relations  (\ref{treerel})
into the $\beta$-functions of the symmetric
theory
\begin{eqnarray}
\beta^{(1)}_{g_1 }& = & \beta^{(1)}_e - {\sw \over \cw } \beta^{(1)}_{M_W} \nonumber\\
\beta^{(1)}_{g_2}& = & \beta^{(1)}_e + {\cw \over \sw} \beta^{(1)}_{M_W} \nonumber\\
\lambda ^{-1}\beta^{(1)}_{\lambda} & = & 2 \Bigl(
 \beta^{(1)}_e + 2 {\cos 2 \theta_W \over \sin 2 \theta_W}
                    \beta^{(1)}_{M_W} + \beta^{(1)}_{m_H} \Bigr) \nonumber\\
\beta^{(1)}_{G_{f_i}} & = & \beta^{(1)}_e + 
2  {\cos 2 \theta_W \over \sin 2 \theta_W}
                    \beta^{(1)}_{M_W} + \beta^{(1)}_{m_{f_i} } .
\label{eq:betrel4}
\end{eqnarray}
These relations can be verified using the explicit form of the 
$\beta$-functions for the spontaneously broken theory given above and
the $\beta$-functions in the symmetric parameterization from
\citere{Ara}.
Similar relations occur if renormalization constants introduced for the
parameters of the symmetric theory are expressed in terms of the
renormalization constants of the electric charge and the particle
masses of the spontaneously broken theory (see e.g.~\cite{bhs}).
The simple relations \refeq{eq:betrel4} between
the $\beta$-functions are not expected to hold beyond one-loop order
since scheme-dependent corrections enter in higher orders.

\newsubsection{QED}

Considering only the charged fermions  and conservation  of electromagnetic
charge one is able to construct usual QED as it is embedded in the
classical action of the Standard Model.
Among the vector bosons it only includes
the photon which  is now an abelian gauge field by construction.
Since
all fermion masses are invariant under QED transformations,
one does not have to introduce a Higgs field into  the theory. Charged scalars
are not added because they are unphysical particles in the Standard Model
and even more their interaction with fermions is only well-defined
if we include the non-abelian symmetries of weak interactions.

The QED-action takes then the usual form
\begin{equation}
\Ga^{QED}_{cl}  = \intd \Bigl( - \frac 14 
F^{\mu\nu} F_{\mu\nu} 
+ \sum_{f_i} \bigl( i\bar f_i \ga^\mu D_\mu f_i - m_{f_i}\bar f_i f_i \bigr)
 - {1\over 2\xi} \left(\partial_{\mu} A^{\mu}\right)^2 \Bigr),
\end{equation}
and 
\begin{equation}
D_\mu f_i = \partial _\mu f_i - i e Q_f A_\mu f_i .
\end{equation}
The action and the charges are defined by the QED Ward identity
\begin{equation}
\label{wiem}
\Bigl( e {\mathbf w}_{em} - \partial^\mu {\delta \over \delta A^\mu} \Bigr) \Ga
= {1 \over \xi} \Box  \partial A .
\end{equation}
The CS equation is given to all orders by:
\begin{equation}
\label{csem}
\Bigl\{ m\partial _m + \beta_e e\partial _e -  \ga_A 
 \Bigl(
{\cal N}_A
 -\xi \partial_\xi \Bigr)
- \sum_{f_i}\ga_{f_i} {\cal N}_{f_i}
 \Bigr\}  \Ga =
\sum_{f_i}m_{f_i} \intd \!
{\delta \Ga \over \delta \hat \varphi_{f_i}} ,
\end{equation}
with
\begin{eqnarray}
m \partial_m & = &
\kappa \partial _\kappa + \sum_{f_i} m_{f_i} \partial m_{f_i} \nonumber \\
{\cal N}_A & = &  \intd\! 
 A^\mu {\delta \over \delta A^\mu} 
\\
{\cal N}_{f_i} & = & \intd\! \bigl( \bar f_i {\delta \over \delta \bar
f_i} - {\delta \over \delta f_i} f_i \bigr) .
\nonumber
\end{eqnarray}
The $\hat \varphi_{f_i}$ are {\it external} scalar fields, which
are introduced for defining
the soft breaking of dilatations.
In pure QED  the $\beta$-function
is related to the anomalous dimension of the photon field
\begin{equation}
\label{abrelqed}
\beta_e = \ga_A .
\end{equation}
The 1-loop contributions to
the  $\beta$-function are exactly the same as the ones which
contribute to the $\beta$-function of the electromagnetic coupling
 in the electroweak Standard Model
from fermions (\ref{eq:betae}):
\begin{equation}
\beta^{(1)}_e = {e^2 \over 4 \pi^2} \frac 13 N_F (Q^ 2_e + 3 Q^ 2_u + 3
 Q^ 2_d ) .
\end{equation}
The Standard Model $\beta$-function in addition includes contributions
from unphysical scalars $\phi^ {\pm}$ and especially 
non-abelian contributions from charged vector bosons 
and charged ghosts with negative sign,
which sum up to a negative sign if one considers only one family.
In QED  one defines   the effective coupling
by the characteristic equation
\begin{equation}
\label{chareq}
{{\partial e} \over \partial t} = e \beta _e  \quad \hbox{with} \quad
t = \ln  \left| \frac{p^2}{\kappa^2} \right| .
\end{equation}
Its solution is interpreted as the momentum and scale dependence
of the interaction strength in the high-energy region. Due to the
relation between the anomalous dimension of the photon field
and the $\beta$-function
(\ref{abrelqed}) the solution can also be identified with the complete
Dyson-summed 
photon propagator.
The behavior of the 1-loop effective
coupling of the Standard Model as solution of the corresponding
characteristic
equation (\ref{chareq})
differs from the
QED-behavior by  the additional non-abelian contributions from ghosts
and
vector bosons. It
approaches zero if one includes only one family, and goes much
more flat to infinity if one takes into account two or three families
of fermions. Moreover, since the abelian relation of QED (\ref{abrelqed}) is
replaced by the relation (\ref{abrel}) an interpretation of the
running  coupling in terms of 2-point
photon Green functions is not clear in the Standard Model.

For evaluating the CS equation  it is important to
control the  $\beta$-functions of higher orders. In particular
it has to be shown that the result which one obtains by integrating
the CS equation is meaningful if one includes only lowest order 
$\beta$-functions. 
In this context
it is important to mention that in pure QED 
there exist also equations for the
differentiation with respect to single fermion masses.
The fermion-mass equations read
\begin{equation}
\label{mfi}
\Bigl(m_{f_i}
\partial _{m_{f_i}} + \beta^{f_i}_e e\partial_e - \ga^{f_i}_A \bigl( 
{\cal N}_A - 2\xi \partial_ \xi \bigr) 
 - \sum_{f_j} \ga_{f_j}^{f_i}
{\cal N}_{f_j} \Bigr) \Ga = m_{f_i}
\intd \!{\delta \over \delta \hat \varphi_{f_i}} \Ga .
\end{equation}
The Ward identity (\ref{wiem})
relates the $\beta$-function and anomalous dimension
of the photon for any of these equations,
\begin{equation}
\beta_e^{f_i} = \ga_A^{f_i} .
\end{equation}
The $\beta$-functions of these equations depend strongly on
the normalization condition imposed for the photon residuum, e.g.
\begin{equation}
\label{photonnorm}
\partial_{p^2} \Ga^T _{AA} \Big|_ {p^2 = \kappa^2} = 1\/ .
\end{equation}
 They all vanish to all orders 
if the residuum of the photon is normalized at a 
normalization
point at infinity ($\kappa^2 \to -\infty$),
\begin{equation}
\lim_{\kappa^2 \to -\infty}\beta_e^{f_i} = 0 \/ .
\end{equation}
 Taking the normalization point at zero momentum
they are given in 1-loop order by
\begin{equation}
\beta_e^{e_i} (\kappa^2 = 0)= {e^2 \over 4 \pi^2} \frac 13 
Q^ 2_e + O(\hbar^2 )
 \qquad \beta_e^{q_i}(\kappa^2 = 0) = {e^2 \over 4 \pi^2} \frac 13  3 Q^ 2_q + O(\hbar^2 ) \, ,
      \quad q = u,d .
\end{equation}
In higher orders  
the consistency equations of the fermion-mass equation (\ref{mfi})
with
the CS equation (\ref{csem}) in QED give important restrictions 
on the mass dependence of the various $\beta$-functions. 
The fermion-mass equations together with the consistency equations
are the main ingredients for 
being able to formulate the running of the effective coupling from the
low-energy to the high-energy region in the simplified version introducing 
step functions.
Eventually they also
 allow to study decoupling of fermions and the construction of
effective low-energy theories. 

Contrary to QED  none of these fermion-mass
differential equations exists in the Standard Model,
since differentiations
with respect to single mass parameters produce hard insertions
of Yukawa interactions. 
As a consequence the CS $\beta$-functions can in principle 
depend on mass ratios in an arbitrary way. The appearance of such 
a logarithmic mass dependence in theories with spontaneously broken symmetry 
has been demonstrated in the simple Higgs-Yukawa-model~\cite{InvCh},
and the respective analysis
has to be continued  to  the Standard Model by a systematic construction
of one-loop induced higher-order contributions (cf.~the discussion at
the end of section 5).

\newsection{Conclusions}
In theories with spontaneously broken symmetry the CS equation plays a
crucial role for a systematic investigation of the large-momentum
behavior of higher-order contributions. It is furthermore an
important instrument within the framework of abstract renormalization 
allowing to determine the independent parameters of the theory in a 
scheme-independent way. In this paper we have derived the CS equation 
for the electroweak Standard Model in the on-shell parameterization and
evaluated all its coefficient functions in one-loop order. 

As a direct application, we have shown that
the ghost mass ratio is an independent parameter of the model. It is
renormalized independently from the vector-boson mass ratio, and
consequently the choice of setting these parameters equal in lowest
order is not stable under renormalization.

We have compared the CS equation of the Standard Model with the ones of the
symmetric $SU(2) \times U(1)$ theory and of QED. While the
one-loop $\beta$-function of the electromagnetic coupling depends only
on the coupling itself and is QED-like in this sense, due to
non-abelian interactions it receives contributions with negative sign,
which dominate over the contributions of the fermions if only one
family of fermions is considered. The one-loop $\beta$-functions in the
on-shell parameterization can be related to the $\beta$-functions of the
symmetric theory in a simple way. These simple relations are not
expected to hold anymore beyond one-loop order, since the higher-order
$\beta$-functions in the on-shell parameterization will contain
logarithms of the masses which are absent in the symmetric theory.

With the CS equation and its one-loop coefficient functions 
we have provided the basic tools necessary for an investigation of
one-loop induced higher-order contributions in the electroweak Standard
Model, as e.g.\ the leading logarithms. Since 
a restricted choice of the gauge fixing will in general not be stable 
under renormalization, we have given the explicit form of all
one-loop coefficient functions in the most general linear gauge
compatible with rigid symmetry transformations. As an example we have
determined the leading quadratic logarithms of the photon self-energy
in 2-loop order. Contrary to QED it is seen that the quadratic logarithms
of the photon self-energy in the asymptotic region are non-vanishing. In 
this context we have also discussed the  possible
sources for the appearance of large mass-dependent
logarithms in 2-loop order. 
 All these contributions strongly depend on the normalization
conditions  imposed for fixing the  free parameters of the Standard
Model. If the Standard Model is renormalized in the on-shell schemes,
these contributions are expected to be present and to depend 
logarithmically on the
different mass ratios.
Due to
the presence of massless  particles on-shell conditions which allow to
diagonalize the mass-matrix of the neutral massive/massless particles
on-shell are 
crucial for obtaining
off-shell infrared-finite Green functions in higher orders.
A systematic analysis of mass-dependent
higher-order contributions is needed for an improvement of the 
perturbative series on the basis of a summation of
large higher-order terms by using the CS or RG equation and its 1-loop 
$\beta$-functions. This issue is the subject of further investigations.
\par

{\it Acknowledgements}
E.\ K.~wants to thank the Institute of Theoretical Physics, Karlsruhe,
 and in particular
Prof.~W.~Hollik for kind hospitality while the final version  of the
 paper has been worked out. We are grateful to K. Sibold for critical
comments and discussions at different stages of the present work.

\vspace{12mm}
\renewcommand{\theequation}{A.\arabic{equation}}
\setcounter{equation}{0}
\begin{flushleft}
{\large \bf Appendix }
\end{flushleft}
\nopagebreak
\medskip
\nopagebreak
In this appendix we give the propagators of the free fields
 in the general linear
gauge defined in (\ref{eq:gf}) and (\ref{Fgen}).  The
propagators of vector and scalar fields are non-diagonal in the 
vector/scalar and in the 
neutral vector fields. 
We omit propagators with
$B_a$~fields, since they do not contribute in loops and 
are not relevant for determining the coefficient
functions of the Callan-Symanzik equation. 
 The free field propagators 
determined from the action with $B_a$ fields
 are equivalent to the ones determined in usual $R_\xi$~gauges
where $B_a$~fields are eliminated via their equations of motion (see 
(\ref{Rxi})).

 For determining the $\beta$-functions and
anomalous dimensions we have taken the choice (\ref{xidef}) --
(\ref{zetadef3}), which is compatible with rigid symmetry. Moreover, 
from boson propagators
only those terms contribute to the CS coefficient functions in 1-loop order
 which behave like
$1 \over p^2$  for asymptotically large $p^2  $, the other terms contribute
as soft mass insertions on the r.h.s.~of the CS equation.

We have taken the following definitions for determining the free field
propagators of vector and scalar fields: 
\begin{equation}
\label{Deltadef}
\sum_l \int d^4 z \Ga^{(0)}_{\varphi_k \varphi_l} (x,z)
\Delta_{\varphi_l \varphi_m} (z,y) = i \delta_{km} \delta^4(x -y) .
\end{equation}
Here $\varphi_k$ denotes all vector and scalar fields of the Standard  Model,
and the index $k$ is understood to include field indices as well as Lorentz
indices: 
\begin{equation}
\varphi_k = (W_{\mu}^+, W_{\mu}^-, Z_{\mu},A_{\mu},
 \phi^+, \phi^-, H, \chi) .
 \end{equation}
The $\Ga^{(0)}_{\varphi_k \varphi_l} $ denote the lowest-order vertex 
functions derived from the generating functional of 1PI Green functions,
\begin{equation}
\Ga^{(0)}_{\varphi_k \varphi_l}(x,y) \equiv 
{\delta^2 \Ga_{cl} \over \delta \varphi_k (x) \delta \varphi_l(y) } \Bigg|_{{
all\;  fields} \; = \; 0} .
\end{equation}
The free field propagators $\Delta_{\varphi_k \varphi_l} (x,y)$ are
the time ordered vacuum expectation values of free fields:
\begin{equation}
 \Delta_{\varphi_k \varphi_l} (x,y) = \big\langle 0 |  T\,  
\varphi_k (x)\, \varphi_l (y)| 0 \big\rangle ^{(0)}.
\end{equation}
The Fourier transformed propagators are defined according to
the  conventions:
\begin{eqnarray}
\label{FT}
 \Delta_{\varphi_k \varphi_l} (x,y)  & = & 
\int \frac { d^4 p} {(2\pi)^4} \Delta_{\varphi_k \varphi_l} (p,- p) 
e ^ {-ip (x -y )}  \\
\label{FTb}
 (2\pi)^4 \delta^4 (p+q)\Delta_{\varphi_k \varphi_l} (p,q) & = & 
\int  { d^4 x} {d^4 y} \Delta_{\varphi_k \varphi_l} (x,y) 
e ^ {i(p x +q y )} .
\end{eqnarray}

\begin{enumerate}
\item Free field propagators of the charged vector and scalar fields \\
Starting from the general gauge-fixing action (\ref{eq:gf}) we find
with the notation
\begin{equation}
\xi_W \equiv \xi_{+-} = \xi_{-+} 
\end{equation}
the following expressions:
\begin{eqnarray}
\Delta_{\phi^+ \phi^-} (p^2) &= &
\frac i{p^2 - \zeta _W M^ 2_W}
 \left( 1- \frac {(\xi_W - \zeta_W  )M_W^ 2}{p^2 - \zeta _W M^ 2_W} \right)
\\
\Delta^L_{\phi^+ W_-} (p^2) &= &  
\frac {i (\xi_W - \zeta_W  )M_W}{( p^2 - \zeta _W M^ 2_W )^ 2}
\\
\Delta^T _{W_+ W_-} (p^2) & = & \frac{i}{p^2 - M_W^2}
\\
\Delta^L _{W_+ W_-} (p^2) & = & 
\frac {i}{p^2 - \zeta _W M^ 2_W}
 \left( \xi_W + \frac {(\xi_W - \zeta_W  )
\zeta_W M_W^ 2}{p^2 - \zeta _W M^ 2_W} \right) .
\end{eqnarray}
Here we have defined the longitudinal and transverse parts of
vector propagators by
\begin{eqnarray}
\label{transvec}
\Delta_{V^\mu_a V^\nu_b} (p , - p) & = &
- \left(\eta^{\mu\nu} - \frac {p^\mu p^\mu} {p^2} \right)
\Delta_{V_a V_b}^T (p^2)
 - \frac {p^\mu p^\mu} {p^2} \Delta_{V_a V_b}^L (p^2) .
\end{eqnarray}
Similarly we have split off the 4-momentum  $p^\mu$ from the scalar/vector
propagator:
\begin{equation}
\label{longscalvec}
\Delta_{\phi_a V^\mu_b}( p, -p) =  p^\mu \Delta^L_{\phi_a V_b} (p^2)  . 
\end{equation}
The remaining propagators are obtained by complex conjugation: 
\begin{equation}
\label{Deltacc}
\Delta^ *_{\varphi_k \varphi_l} (p,- p) = - 
\tilde I_{k k'} \tilde I_{l l'}\Delta_{\varphi_{k'} \varphi_{l'}} (- p, p) .
\end{equation}
The matrix $\tilde I $ is defined in analogy to eq.~(\ref{tildeI}),
and (\ref{Deltacc}) means in particular:
\begin{equation}
\Delta^ *_{\phi^+ W^\mu_-} (p,- p) = -  \Delta_{\phi^- W^\mu_+} (-p, p) 
= \Delta_{\phi^- W^\mu_+} (p, - p) .
\end{equation}

\item Free field propagators of the neutral scalar and vector fields\\
With the notation 
\begin{equation}
\xi_Z \equiv \xi_{ZZ}  \qquad \qquad 
\xi_A \equiv \xi_{AA}  
\end{equation} 
for the arbitrary gauge parameters of (\ref{eq:gf}) one obtains
\begin{eqnarray}
\Delta_{HH} (p^2) &= &
\frac i{p^2 - m_H^ 2}
 \\
\Delta_{\chi \chi} (p^2) &= &
\frac i{p^2 - \zeta _Z M^ 2_Z}
 \left( 1- \frac {(\xi_Z - \zeta_Z  )M_Z^ 2}{p^2 - \zeta _Z M^ 2_Z} \right)
\\
\Delta^L_{\chi Z} (p^2) &= &  
\frac {- (\xi_Z - \zeta_Z  )M_Z}{( p^2 - \zeta _Z M^ 2_Z )^ 2}
\\
\Delta^L_{\chi A} (p^2) &= &  
\frac { M_Z}{p^2( p^2 - \zeta _Z M^ 2_Z )}
\left(\zeta_A   - \xi_{AZ}  
 - \frac{(\xi_Z - \zeta_Z) \zeta_A M_Z^2}{p^2 -\zeta_Z M_Z^ 2} \right)
\\
\Delta^T _{ZZ} (p^2) & = & \frac{i}{p^2 - M_Z^2}
\\
\Delta^T _{ZA} (p^2) & = & 0
\\
\Delta^T _{AA} (p^2) & = & \frac{i}{p^2 }
\\
\Delta^L _{ZZ} (p^2) & = & 
\frac {i}{p^2 - \zeta _Z M^ 2_Z}
 \left( \xi_Z + \frac {(\xi_Z - \zeta_Z  )
\zeta_Z M_Z^ 2}{p^2 - \zeta _Z M^ 2_Z} \right)
\\
\Delta^L _{ZA} (p^2) & = & 
\frac {i}{p^2 - \zeta _Z M^ 2_Z}
 \left( \xi_{ZA} + \frac {(\xi_Z - \zeta_Z  )
\zeta_A M_Z^ 2}{p^2 - \zeta _Z M^ 2_Z} \right) \\
\Delta^L _{AA} (p^2) & = & 
\frac {i}{p^2 }
 \left( \xi_A + \frac {(2\xi_{AZ} - \zeta_A  )
\zeta_A M_Z^ 2}{p^2 - \zeta _Z M^ 2_Z} 
+ \frac {(\xi_{Z} - \zeta_Z  )
\zeta_A M_Z^ 4}{(p^2 - \zeta _Z M^ 2_Z)^ 2} 
\right) .
\end{eqnarray}
Non-diagonal propagators that are not given in the above list vanish 
identically because of CP-invariance of the free field action.

\item Free field propagators of the Faddeev-Popov fields \\
The free field propagators of the Faddeev-Popov fields are derived from
the bilinear part of the ghost action (\ref{ghostdiag}).
They are diagonal according 
to the construction outlined in section 3, eqs.~(\ref{gaugepar}) -- 
(\ref{STG}). For this reason they have their conventional form:
\begin{eqnarray}
\Delta_{ c_+ \bar c_-} (p^2) & = & \frac{i} {p^2 - \zeta_W M_W^ 2} \\
\Delta_{ c_Z \bar c_Z} (p^2)& = & \frac{i} {p^2 - \zeta_Z M_Z^ 2} \\
\Delta_{ c_A \bar c_A} (p^2)& = & \frac{i} {p^2 } \\
\Delta_{ c_A \bar c_Z} (p^2) & = & \Delta_{ c_Z \bar c_A} (p^2) \hspace{3mm} =
\hspace{3mm}  0 .
\end{eqnarray}
They are derived from the classical action in an equivalent way to
(\ref{Deltadef}),
\begin{equation}
\sum_d\int d^4 z \Ga^{(0)}_{ c_a \bar c_d}(x,z) \Delta_{c_b \bar c_d} (y,z)
= i \delta_{ab} \delta^4 (x -y) ,
\qquad
\Ga^{(0)}_{ c_a \bar c_b}(x,y) \equiv {\delta^2 \Ga_{cl}
 \over \delta c_a(x) \bar c_b(y)}  ,
\end{equation}
and are related to the time ordered vacuum expectation values of free fields by
\begin{equation}
\Delta_{ c_a \bar c_b} (x,y)
= \big\langle 0|T\, c_a(x) \,\bar c_b (y)| 0\big \rangle ^{(0)}.
\end{equation}
Fourier transformation is defined as in (\ref{FT}), (\ref{FTb}).

\item Free field propagators of fermions \\
For completeness we also give the free field  propagator of a Dirac fermion:
\begin{equation}
\Delta_{f \bar f}( p, - p) =  \frac{i(\, {\! \! \not \! p \,}
 + m_{f})}{p^{2}-m_{f}^{2}} .
\end{equation}
It is  determined from the classical action by 
\begin{equation}
\sum_\beta \int d^4 z \Ga^{(0)}_{ \bar f_\a  f_\beta}(x,z) \Delta_{f_\beta
 \bar f_\ga} (z,y)
= i \delta_{\a\ga} \delta^4 (x -y)
\end{equation}
with
\begin{equation}
\Ga^{(0)}_{ \bar f  f}(x,y) \equiv {\stackrel{\rightarrow}{\delta}
 \over \delta \bar f
(x)} \Ga_{cl}
{\stackrel{\leftarrow}{\delta} \over \delta f (y)}.
\end{equation}
Differentiation with respect to the adjoint spinor 
$\bar f$ is applied from the left,
whereas differentiation with respect to the spinor $ f$ 
is applied from the right, $\a, \beta, \ga$ are spinor indices.
The free field propagator is related to the time ordered vacuum expectation
value of free fields by
\begin{equation}
\Delta_{  f \bar f} (x,y)
= \big\langle 0|T\, f(x) \,\bar f  (y)| 0\big \rangle ^{(0)}.
\end{equation}
Fourier transformation is defined as in (\ref{FT}), (\ref{FTb}).

\end{enumerate}

\end{document}